\documentclass[aps,prc,twocolumn,superscriptaddress]{revtex4-2}

\usepackage{amsmath, mathtools}
\usepackage[dvipdfmx]{graphicx}
\usepackage{url}
\usepackage{color}
\begin{document}


\title{Photodisintegration Cross Section of $^4$He in the Giant Dipole Resonance Energy Region}



\author{M.~Murata}
\affiliation{Research Center for Nuclear Physics (RCNP), Osaka University, Ibaraki, Osaka 567-0047, Japan}
\author{T.~Kawabata}
\affiliation{Department of Physics, Osaka University, Toyonaka, Osaka 560-0043, Japan}
\author{S.~Adachi}
\affiliation{Cyclotron and Radioisotope Center (CYRIC), Tohoku University, Sendai, Miyagi 980-8578, Japan}
\author{H.~Akimune}
\affiliation{Faculty of Science and Engineering, Konan University, Kobe, Hyogo 658-8501, Japan}
\author{S.~Amano}
\affiliation{Laboratory of Advanced Science and Technology for Industry (LASTI), University of Hyogo, Ako, Hyogo 678-1205, Japan}
\author{Y.~Fujikawa}
\affiliation{Department of Physics, Kyoto University, Sakyo, Kyoto 606-8502, Japan}
\author{T.~Furuno}
\affiliation{Department of Physics, Osaka University, Toyonaka, Osaka 560-0043, Japan}
\author{K.~Inaba}
\affiliation{Department of Physics, Kyoto University, Sakyo, Kyoto 606-8502, Japan}
\author{Y.~Ishii}
\affiliation{Department of Physics, Kyoto University, Sakyo, Kyoto 606-8502, Japan}
\author{S.~Miyamoto}
\affiliation{Laboratory of Advanced Science and Technology for Industry (LASTI), University of Hyogo, Ako, Hyogo 678-1205, Japan}
\affiliation{Institute of Laser Engeneering, Osaka University, Suita, Osaka 565-0871, Japan}
\author{M.~Tsumura}
\affiliation{Department of Physics, Kyoto University, Sakyo, Kyoto 606-8502, Japan}
\author{H.~Utsunomiya}
\affiliation{Faculty of Science and Engineering, Konan University, Kobe, Hyogo 658-8501, Japan}
\affiliation{Shanghai Advanced Research Institute, Chinese Academy of Science, Shanghai 201210, China}


\date{\today}

\begin{abstract}
We simultaneously measured
the $^4$He($\gamma$, $n$)$^3$He and $^4$He($\gamma$, $p$)$^3$H reactions 
in the energy range around the giant dipole resonance.
A quasi-mono-energetic photon beam produced via the laser Compton scattering technique 
was irradiated on the active-target time-projection chamber filled with helium gas,
and trajectories of charged decay particles emitted from $^4$He were measured.
Our data suggest that the $^4$He($\gamma$, $n$)$^3$He and $^4$He($\gamma$, $p$)$^3$H cross sections
peak around 26~MeV.
This result contradicts the previous experimental data reported by Shima \textit{et al.}
but is consistent with other experimental results. 
\end{abstract}


\maketitle

\section{Introduction}
\label{sec_intro}
The isovector giant dipole resonance (GDR) is one of the most examined nuclear resonances,
which is excited through the $E1$ transition of nuclear ground states.
It is a representative example of the collective excitation modes of atomic nuclei in which protons and neutrons coherently oscillate in antiphase,
and its energy-integrated cross section exhausts approximately 100\% of the Thomas-Reiche-Kuhn sum-rule value~\cite{Bohr1969}.
The photodisintegration reaction is a suitable probe to investigate the GDR
because the photoabsorption dominantly induces  $E1$ transition 
as the long wave-length approximation commonly holds in various nuclei.

The cross section of the $^4$He photodisintegration reaction in the GDR energy region has recently attracted research interest because it is an important aspect
for understanding nucleosynthesis process in the universe.
One example is the $\nu$-process in the He-layer of core-collapse supernovae \cite{Woosley1990}. 
In this process, rare elements such as $^7$Li and $^{11}$B are produced through a series of nuclear reactions 
initiated by the $^4$He($\nu$,$\nu ^\prime n$) and $^4$He($\nu$,$\nu ^\prime p$) reactions~\cite{Suzuki2013}.
The giant resonances, such as GDR and spin-dipole resonances, make a dominant contribution on these reactions 
because the $^4$He($\nu$,$\nu^{\prime}$) reaction primarily excites these resonances with $L=1$ ~\cite{Nakamura2017}.
The Gamow-Teller resonance is approximately forbidden in $^4$He due to the system's double magicity.
Because of technical difficulties in measuring neutrino-nucleus reactions,
estimation of neutrino-nucleus reaction cross sections for examining the $\nu$-process relies on the nuclear structure theories.
Nuclear structural information is therefore a source of uncertainty with regard to the calculations.
By using the analogy between electromagnetic responses and weak responses of nuclei~\cite{Ejiri2013},
the experimental cross sections of the $^4$He photodisintegration reaction can provide
a criterion to test the validity of the estimated neutrino-nucleus reactions cross sections.

Furthermore, a relationship was also discussed between the $^4$He photodisintegration and the lithium problem in the Big Bang nucleosynthesis (BBN) ~\cite{Kusakabe2008}.
The lithium problem is an unsolved discrepancy between the primordial abundances of lithium isotopes 
estimated from the astronomical observation and those predicted from the BBN calculation.     
It is an intriguing problem that the primordial abundance of $^7$Li estimated from observing the metal-poor halo starts
is deficient with regard to the standard BBN prediction by a factor of three~\cite{Ryan2000, Melendez2004}. 
In addition, the possibility of an overabundance of $^6$Li at the level of approximately three orders of magnitude was also suggested from the spectroscopic measurements \cite{Asplund2006}.
It was proposed that the extended BBN modified with non-thermal photons produced through radiative decays of unstable relic neutral massive particles
might solve the lithium problem~\cite{Kusakabe2008}.
Although the relic particle has not yet been observed, its possible mass, abundance, and lifetime are constrained by the cross sections of the $^4$He photodisintegration reaction.

The experimental studies on the $^4$He photodisintegration reaction have continuously been reported 
since 1950s~\cite{Gorbunov1958_1,Gorbunov1958_2,Gardner1962,Gemmell1962,Clerc1965,Gorbunov1968,
Meyerhof1970, Sanada1970, Berman1972, Irish1973, Malcom1973, Arkatov1974, Balestra1977,Berman1980, Ward1981, McBroom1982, 
Calarco1983, Bernabei1988, Feldman1990, Komer1993, Hoorebeke1993, Hahn1995, Shima2005, Nilsson2007,Nakayama2007, Shima2010, Tornow2012, Raut2012}.
Most of the previous studies measured one of the $(\gamma, n)$ and $(\gamma, p)$ reactions
and fewer studies were conducted using the simultaneous measurement~ 
\cite{Gorbunov1958_1,Gorbunov1958_2,Gorbunov1968,
Arkatov1974, Balestra1977,Bernabei1988, Shima2005, Shima2010, Tornow2012,Raut2012}.
Some authors measured the photodisintegration reactions 
directly~\cite{Gorbunov1958_1,Gorbunov1958_2,Clerc1965,Gorbunov1968,
Sanada1970, Berman1972, Irish1973, Malcom1973, Arkatov1974, Balestra1977,Berman1980, 
Bernabei1988, Hoorebeke1993, Shima2005, Nilsson2007,Shima2010, Tornow2012,Raut2012},
while others deduced its cross section from the surrogate reactions, such as the radiative capture reactions, which is the inverse reaction of the photodisintegration reactions
\cite{Gardner1962,Gemmell1962,Meyerhof1970,Ward1981, McBroom1982, Calarco1983, Feldman1990, Komer1993, Hahn1995, Nakayama2007}.
Although the direct measurements in early years were conducted with continuous-energy photon beams generated with bremsstrahlung,
the quasi-mono-energetic beam generated from the laser Compton scattering (LCS)~\cite{Bernabei1988, Shima2005,Shima2010, Tornow2012,Raut2012}
or tagged photon technique~\cite{Nilsson2007} were employed in the recent measurement.

However, these previous results exhibit significant deviation beyond their respective error ranges.
Therefore, it is difficult to extract reliable information to constrain theoretical calculations.
Here, we focused on the recent experimental studies~\cite{Shima2005, Shima2010, Raut2012, Tornow2012}
in which quasi-mono-energetic photon beams were employed, and both the $(\gamma, n)$ and $(\gamma, p)$ channels were simultaneously measured. 
Regarding the reaction probe, quasi-mono-energetic photon beams are more suitable than continuous-energy beams
since the unfolding procedure is necessary for the continuous-energy beams. 
This unfolding process could cause significant errors on the cross sections,
unless the stability of the accelerator, sufficient counting statistics, and understanding of the beam-energy spectrum were guaranteed.
Furthermore, the simultaneous measurement is desirable to obtain reliable cross sections of the $(\gamma, n)$ and $(\gamma, p)$ reactions
because one can cancel out the experimental errors due to the target thickness and the beam flux.

Firstly, Shima \textit{et al.}~\cite{Shima2005} performed the measurement using a time-projection chamber (TPC).
By operating the TPC as an active target, 
they measured the trajectories of charged particles and their energy deposits along the trajectories.
Their deliberate experimental design and careful treatments of the data made it possible to clearly distinguish photodisintegration events from background events.
They reported that no peak structures were observed in the photon-energy dependence of the cross sections 
of the $(\gamma, n)$ and $(\gamma, p)$ reactions at $E_{\gamma} < $ 30 MeV.
A follow-up measurement by Shima \textit{et al.} presented a preliminary result suggesting that the peak of the GDR is located around $E_{\gamma} = $ 32 MeV~\cite{Shima2010}.

Raut \textit{et al.}~\cite{Raut2012} and Tornow \textit{et al.}~\cite{Tornow2012} carried out the measurement 
in the HI$\gamma$S facility using a high-pressure $^4$He-Xe gas scintillator as an active target. 
They identified photodisintegration events of $^4$He 
by measuring the total energy deposit of decay charged particles in the detector.
The thick target and intense beam allowed them to obtain statistically high precision data.
The theoretical calculations ~\cite{Quaglioni2004} describe the measured cross sections reasonably well
and suggest that the peak of the GDR is located around $E_{\gamma} = 26$~MeV.
However, they did not report the cross sections of the $(\gamma, n)$ reactions in the lower energy region of $E_{\gamma} < $ 27~MeV.
Thus, we can not conclude the energy dependence of the cross sections of the $(\gamma, n)$ and $(\gamma, p)$ channels solely from their result.

From a theoretical perspective, an elaborate calculation
using a realistic nuclear force with a three-body interaction was reported~\cite{Horiuchi2012}.
The cross sections of the $^4$He$(\gamma, n)$$^3$He and the $^4$He$(\gamma, p)$$^3$H reactions 
were calculated in the microscopic $R$-matrix approach,
in which final state interactions and two- and three-body decay channels were considered.
Both theoretical cross sections of the $(\gamma, n)$ and $(\gamma, p)$ reactions rise sharply from the threshold and 
become maximal at $E_{\gamma}=$ 26 MeV.
This result is consistent with the cross sections previously calculated using the Lorentz integral transform method~\cite{Quaglioni2004} and 
discrepant with the results by Shima {\it et al.}

The data from Shima {\it et al.} contradicts most of previous experimental data and theoretical calculations,
and thus, the results appear unfavorable.
However, their measurements using an active target TPC to track the charged particle trajectories in the final state are unprecedentedly sophisticated.
Background events could be effectively rejected with information from the particle trajectories.
Therefore, the possibility that their results differing from the other experiments are correct 
cannot be ruled out owing to their improved methodology.
In order to solve this situation, 
we simultaneously measured the cross sections of the $^4$He($\gamma$, $n$)$^3$He and $^4$He($\gamma$, $p$)$^3$H reactions using one experimental setup
with a new active target TPC for elucidating the energy dependence of the cross sections in the GDR region.
In this paper, we report the cross sections of the $^4$He($\gamma$, $n$)$^3$He and $^4$He($\gamma$, $p$)$^3$H reactions 
in the energy region between $E_{\gamma}$ = 23--30~MeV 
measured using a quasi-mono-energetic photon beams generated with the LCS technique
and the active target TPC.

\section{Experiment}
\begin{figure*}
    \includegraphics[width=160mm]{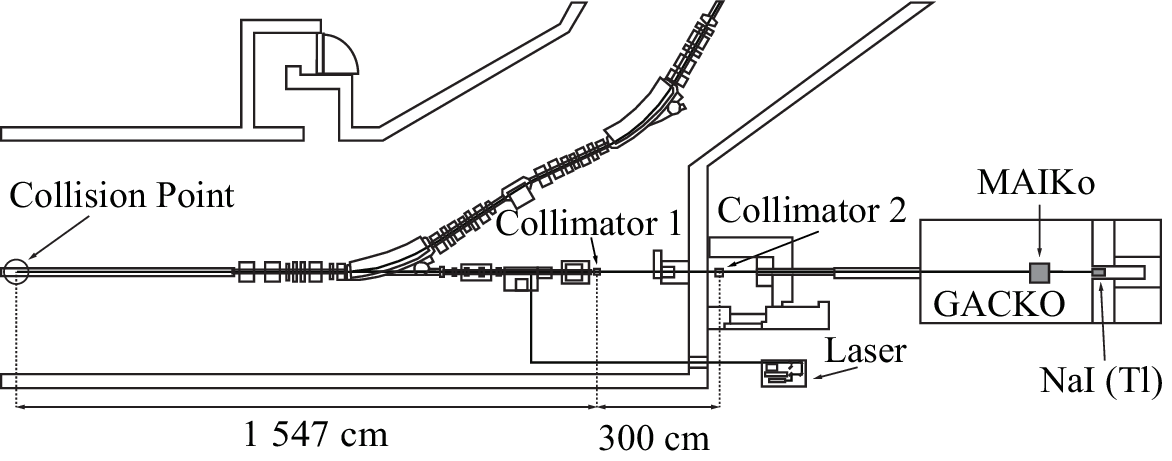}
    \caption{
    Top view of the BL01 at the NewSUBARU synchrotron radiation facility.
    Infrared photon beams generated with the Nd:YVO$_4$ laser modules were guided into 
    the electron storage ring after two reflections.
    The back-scattered photons resulting from the head-on collision with relativistic electrons coming from the left were guided straight to the GACKO beam hutch.
    The MAIKo active target and the NaI(Tl) beam monitor were installed in the GACKO beam hutch.
    On the way to the hutch, photons were sieved with the two collimators to be quasi-mono-energetic photon beams.
    }
    \label{fig_new_subaru}
\end{figure*}

The experiment was performed at the beam line 01 (BL01)~\cite{Miyamoto2006} in the NewSUBARU synchrotron radiation facility.
Figure~\ref{fig_new_subaru} shows an overview of the experimental setup.
A quasi-mono-energetic gamma-ray beam with a maximum energy of $E_{\gamma}$= 23--30 MeV was
produced using the LCS technique.
Laser photons generated using the solid-state laser (Nd:YVO$_4$ $\lambda$ = 1,064 nm) were injected into the storage ring at NewSUBARU and 
impinged on the electron beam circulating in it.
The energy of the laser photon was amplified through Compton scattering with a high-energy electron.
The resultant energy was determined from its scattering angle and the energy of the electron beam.
Photons scattered at $180^\circ$, which had the highest energy, were selected by limiting the scattering angle with the lead collimators installed on the BL01.
A pair of collimators was located at 1,547 and 1,848 cm away from the collision point,
and their apertures were 3 mm and 2 mm, respectively. 

The energy of the gamma-ray beam was tunable by altering the energy of the electron beam, 
and its absolute value was precisely calibrated~\cite{Utsunomiya2014}.
For this experiment, an electron beam with an energy of 1 GeV delivered from the injector linear accelerator was accelerated to 1.146--1.311~GeV in the storage ring.

A linearly polarized gamma-ray beam was employed in the present measurement.
The laser photons were almost 100\% linearly polarized, 
and a half-wave plate was used to tilt the polarization axis of the laser photons at $10^\circ$ from the horizontal axis.
Because the polarization of the LCS gamma rays is maximum at the scattering angle of $180^\circ$~\cite{D'angelo2000},
highly polarized beams were obtained at the GACKO beam hutch.

The gamma-ray beam was monitored with the NaI(Tl) scintillation detector installed at the end of the BL01.
A crystal of the NaI(Tl) scintillator was cut into a cylindrical shape with a diameter of 203.2 mm (8 inches) and a thickness of 304.8 mm (12 inches).
The gamma-ray beam provided through the collimator of the finite-size aperture inevitably had an energy spread
as the energy of the scattered photon correlates to the Compton scattering angle. 
In order to estimate the total photon number irradiated on the target,
the energy spectra deposited to the NaI(Tl) scintillator by the LCS photon beams were acquired during the measurement.
In addition, the energy spectrum with a low-intensity beam was also measured at each beam energy to deduce the energy profile.

\begin{figure}[bp]
    \includegraphics[width=86mm]{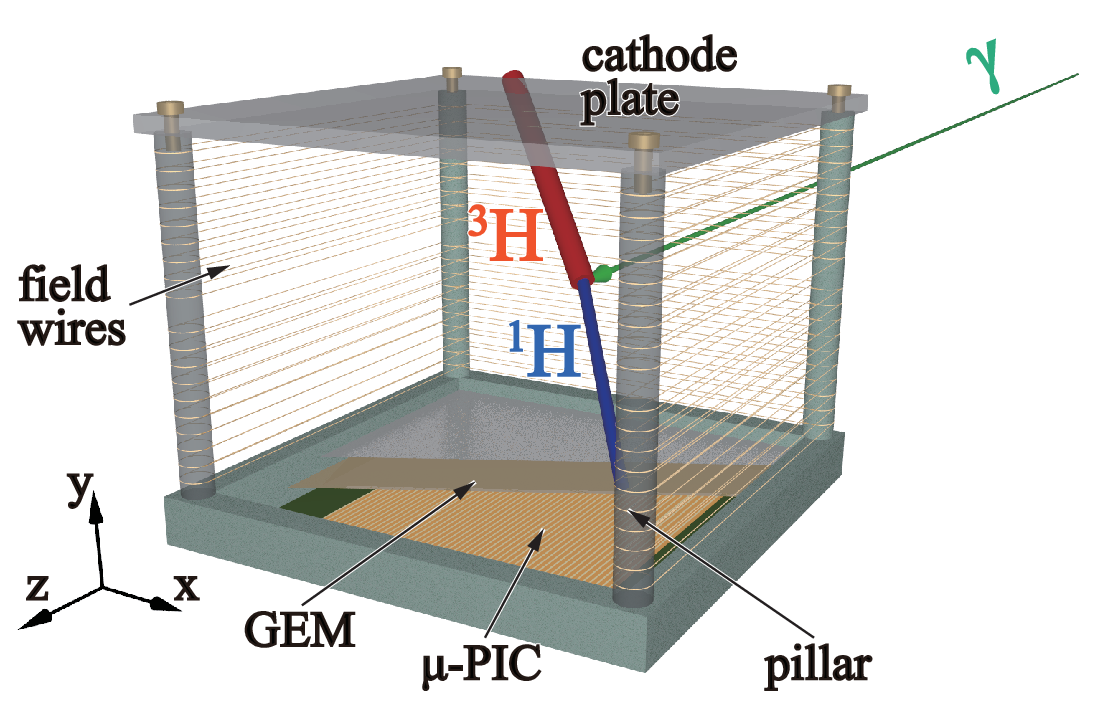}
    \caption{Schematic picture of the MAIKo active target.
    This figure displays a typical $^4$He($\gamma$, $p$)$^3$H event,
    in which a $^4$He nucleus absorbs a photon (thin green arrow emanating from behind the MAIKo active target) 
    to break up into a $^3$H nucleus (the thickest red line) and a proton (thicker blue line).
    Electrons generated along the trajectories of the decay particles were drifted with the uniform electric field to the $\mu$-PIC at the bottom.
    For good visibility, field wires on the front side and the lower half of the GEM are not drawn.
    }
    \label{fig_maiko}
\end{figure}

Charged particles emitted from the photodisintegration of $^4$He were measured using the MAIKo active target~\cite{Furuno2018}.
As shown in Fig.~\ref{fig_new_subaru}, the MAIKo active target was installed in the GACKO beam hutch on the BL01.
A schematic view of the MAIKo active target is shown in Fig.~\ref{fig_maiko}.
The MAIKo active target worked as a TPC.
A negative high voltage was applied on the cathode plate at the top of the MAIKo to form a vertical electric field.
Field wires made of beryllium copper were doubly wound around the pillars with 5-mm intervals to form a uniform electric field.
The sensitive volume of the TPC was the cubic region enclosed in the field cage.
Its dimension defined with the area of the read-out electrodes, and the field cage was 10 $\times$ 10 $\times$ 11 cm$^3$ in width, depth, and height, respectively.
The field cage was aligned so that one dimension of it was normal to the beam direction.
As shown in Fig.~\ref{fig_maiko}, hereinafter we define the three-dimensional Cartesian coordinates such that the $z$-axis is parallel to the beam direction
and the $y$-axis is vertical.

The whole structure shown in Fig.~\ref{fig_maiko} was installed in the vacuum chamber filled with the detection gas.
The detection gas comprised helium with 90\% and CH$_{4}$ with 10\% by their partial pressures, 
in which helium and CH$_4$ served as the target and quench gas, respectively.
The gas pressure was optimized to realize the condition that 
the ranges of the proton and $^3$H were long enough to escape from the sensitive volume, 
whereas that of $^3$He was short enough to stop inside the sensitive volume, as summarized in Table~\ref{tab_measurement}.

As an incident gamma ray was absorbed by a $^4$He nucleus inside the sensitive volume, 
the $^4$He nucleus decayed into a proton and a $^3$H nucleus (a neutron and a $^3$He nucleus), which traveled in the opposite directions.    
Those charged decay particles ionized gas molecules along their trajectories and kicked out electrons.
These electrons were drifted toward the negative direction of the $y$-axis
by the uniform electric field inside the drift cage and gas-amplified by the gas electron multiplier (GEM). 
Finally, the amplified electrons reached the micro-pixel chamber ($\mu$-PIC)~\cite{Ochi2002} at the bottom of the MAIKo active target.
The $\mu$-PIC consisted of 400-$\mu$m pitched 256 $\times$ 256 circular pixels aligned in the square region.
Each pixel was composed of a cylindrical anode electrode with a diameter of 50~$\mu$m at the center and a cathode electrode on the circumference with a diameter of 256~$\mu$m. 
The electrons reached the pixels, caused a Townsend avalanche in the high electric field formed around the anode electrodes,
and induced electric signals on the anode and cathode electrodes.
The anode (cathode) electrodes on the pixels in line were electrically connected to form 256 anode (cathode) strips for signal readout.

\begin{figure}[tbp]
    \includegraphics[width=86mm]{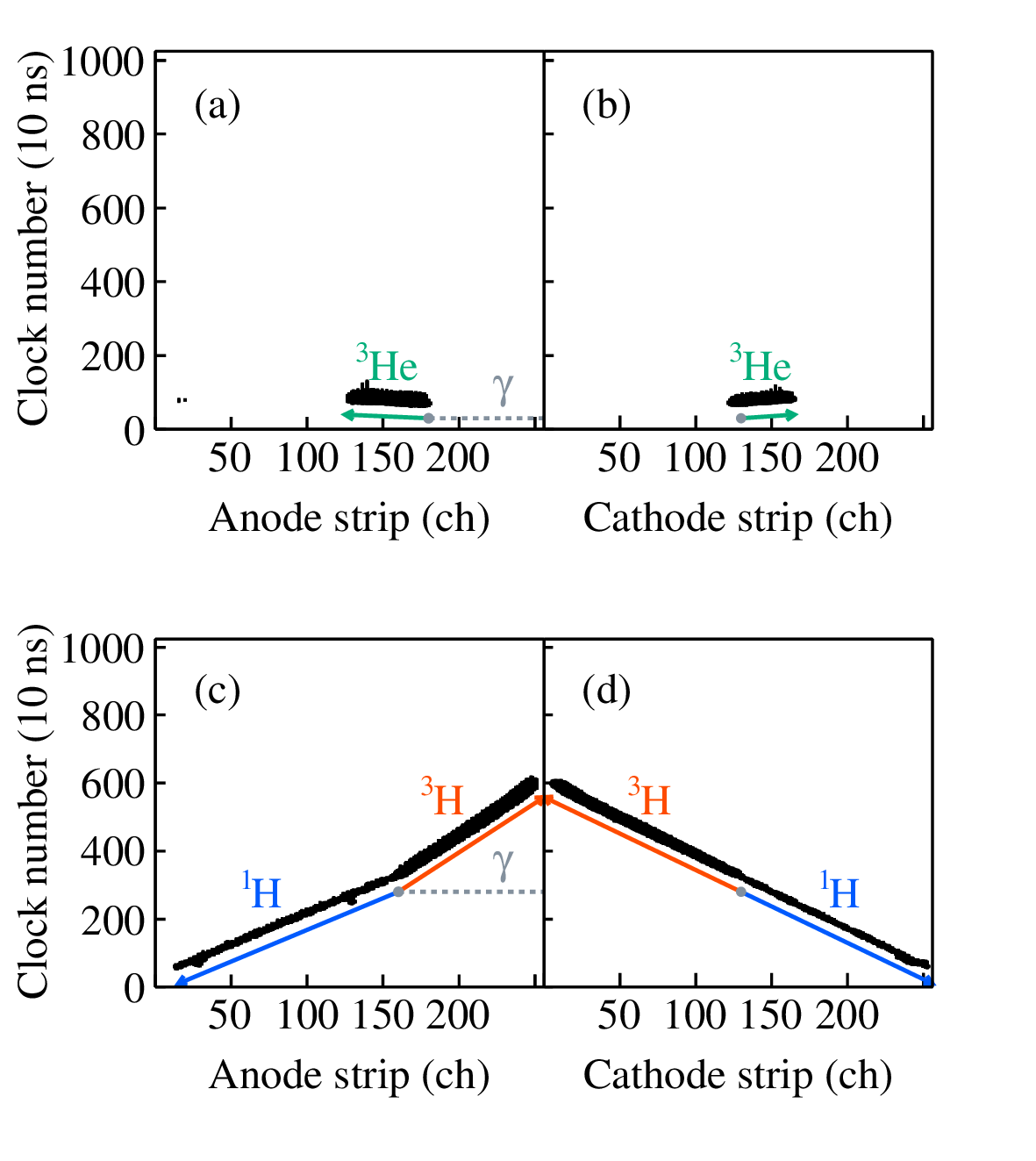}
    \caption{Typical track images measured with the MAIKo active target.
    (a) and (b) present a $^4$He($\gamma, n$)$^3$He event,
    and (c) and (d) indicate a $^4$He($\gamma, p$)$^3$H event.}
    \label{fig_track}
\end{figure}

The electric signals induced on the $\mu$-PIC were read out and processed with the dedicated electronics
with the amplifier-shaper-discriminator (ASD)~\cite{Mizumoto2015}.
When a trigger signal was provided to the signal readout electronics,
the status of the discriminators at every 10 ns (100 MHz) was recorded as a function of the clock number within a time window of 10.24 $\mu$s.
Thus, the acquired data were equivalent to two black-and-white images with 256 $\times$ 1024 pixels
in which the time-over-threshold of the signal was presented with the filled pixels, as shown in Fig.~\ref{fig_track}.
The clock number corresponds to the vertical position from where the trigger event occurred.
In contrast, the strip number is the horizontal position. 
Because the directions of the cathode strips were parallel to the $z$-axis, and those of the anode strips were parallel to the $x$-axis,
the black-and-white image obtained from the anode (cathode) strips presented particle trajectories projected onto the $xy$ ($zy$) plane.
Figures~\ref{fig_track}(a) and (b) show two-dimensional images of a typical $^4$He($\gamma, n$)$^3$He event,
whereas Figs.~\ref{fig_track}(c) and (d) are those of a typical $^4$He($\gamma, p$)$^3$H event.
$^3$He was observed as one thick trajectory,
and proton and $^3$H left two thin and medium long trajectories.
All trajectories of charged particles started from the horizontal center of cathode images,
where the beam-injection point was located.
By combining a couple of images, we could reconstruct the three-dimensional configuration of the charged particle trajectories.

Analog signals on adjacent 32 consecutive strips were summed up,
and the waveform of the summed signal was also sampled at a frequency of 25~MHz.
Because the time constant of the shaping amplifiers was shorter than the typical time scale of the induced signal,
one could obtain energy-deposit information by integrating the waveform.
The summed signals were also employed for the trigger decision.

The trigger signal was generated when the pulse height of any summed signals exceeded the threshold level. 
The threshold level was set sufficiently low so as to avoid missing the photodisintegration events.
The trigger rate was less than 100 Hz, and the data acquisition efficiency was higher than 99\%.

\begin{table}
\caption{\label{tab_measurement} Summary of the pressure and effective thickness of the detection gas. 
$k$ is the index of the beam energy. 
$E_{e}$ and $E_{\mathrm{max}}$ stand for the energy of the electron beam and the maximum energy of the gamma-ray beam, respectively.
Effective thickness is a mass thickness at the depth of the sensitive volume (10 cm). 
}
\begin{ruledtabular}
\begin{tabular}{l c c c c}
    $k$ & $E_{e}$ & $E_{\mathrm{max}}$  & Pressure  & Thickness \\
     & (GeV) & (MeV) & (hPa) &  (mg/cm$^2$) \\
    \hline
    1 & 1.146 & 23.0 & 500 & 1.05 \\
    2 & 1.171 & 24.0 & 500 & 1.05 \\
    3 & 1.195 & 25.0 & 1000 & 2.09 \\
    4 & 1.243 & 27.0 & 1000 & 2.09 \\
    5 & 1.266 &28.0 & 1000 & 2.09 \\
    6 & 1.311 & 30.0 & 2000 & 4.18
\end{tabular}
\end{ruledtabular}
\end{table}

\section{Analysis}
\subsection{Event Selection}

Firstly, candidates of the $^4$He photodisintegration events were selected out of all the events acquired with the MAIKo active target.
The main sources of background events were electrons emitted from Compton scattering and discharges that occurred on the $\mu$-PIC.
Photodisintegration events of $^{12}$C in CH$_4$ molecules were also a source of the background.
Another source would be the photodisintegration events occurring outside the sensitive volume.
The shapes of the analog signals resulting from the Compton scattering events and the discharging events were 
distinctively different from those of the photodisintegration events. 
Furthermore, the photodisintegration events that arose outside the sensitive volume
did not produce any signals significantly higher than the noise level on the $\mu$-PIC electrodes adjacent to the beam axis.
Thus, these background events were safely excluded using the pulse-shape information.
These event selection criteria were set loose enough so as not to exclude the $^4$He photodisintegration events.
After this selection procedure, the number of events was reduced by one order of magnitude.

Secondly, we performed tracking analysis to extract the $^4$He($\gamma, n$)$^3$He and $^4$He($\gamma, p$)$^3$H events 
from the candidate events selected by the pulse-shape analysis.
The $^4$He($\gamma, n$)$^3$He events exhibited one short thick trajectory due to $^3$He, 
which terminated inside the sensitive volume.
On the other hand, the $^4$He($\gamma, p$)$^3$H events left two long thin and medium trajectories 
due to a proton and a $^3$H nucleus which continued to the boundary of the sensitive volume.

The procedure of the tracking analysis was as follows.
Following the cartesian coordinate defined in Fig.~\ref{fig_maiko},
we refer to the track image obtained from the anode (cathode) strips as ZY (XY) image hereafter.

\begin{enumerate}
    \item Clear false hit pixels due to electric noise in the ZY and XY images.
    \item Determine the reaction point in the XY image. The $x$ position of the reaction point was fixed at the center of the beam axis, and the $y$ position was determined by averaging the $y$ positions of the hits around the beam axis.
    \item Track the trajectories starting from the reaction point in the XY image and derive the number, direction, and length of the trajectories.
    \item By using the $y$ position of the reaction point in the XY image, determine the reaction point in 
    the ZY image, and track the trajectories in the ZY image.    
    \item Based on the $y$ positions of the end points in the XY and ZY images, determine the combination of the trajectories in the two images to reconstruct the three-dimensional trajectories.    
\end{enumerate}

After the tracking analysis, the analyzed events were classified with the number of reconstructed trajectories. 
When only one trajectory was reconstructed, this event was classified as a candidate of the $^4$He($\gamma, n$)$^3$He event.
If two trajectories were reconstructed, this event was classified as a candidate of the $^4$He($\gamma, p$)$^3$H event.
For further analysis, we calculated total energy loss in the sensitive volume ($E$) and the differential energy loss ($dE/dx$) of the reconstructed trajectory.
We utilized them to reject the remaining background due to the $^{12}$C photodisintegration.

For the $^4$He($\gamma, n$)$^3$He candidates, we required the five conditions below.
\begin{itemize}
    \item The number of reconstructed trajectories is one.
    \item The reaction point is at least 1 cm distant from the edge of the sensitive volume.
    \item The end point of the trajectory is inside the sensitive volume. 
    \item The correlation between $E$ and the trajectory length is consistent with that of $^3$He.
    \item Kinematically reconstructed beam energy is in the range between $E_{\mathrm{max}}-1$ and $E_{\mathrm{max}}$ MeV.
\end{itemize}
For the $^4$He($\gamma, p$)$^3$H candidates, we imposed five conditions below.
\begin{itemize}
    \item The number of reconstructed trajectories is two.
    \item The reaction point is at least 1 cm distant from the edge of the sensitive volume.
    \item The two trajectories reach the edge of the sensitive volume.
    \item $dE/dx$ of one trajectory is consistent with that of a proton, and $dE/dx$ of the other trajectory is consistent with that of $^3$H.
    \item The two trajectories are oriented in a back-to-back direction in the XY image.
\end{itemize}

Finally, we defined events that satisfy the above conditions the procedure as true photodisintegration events.
Because there should be some true events lost in the tracking and event-selection procedures,
it was necessary to consider the detection efficiency of the true events with the MAIKo active target to determine the cross sections.

\subsection{Efficiency Estimation}
The detection efficiency with the MAIKo active target was estimated using a Monte Carlo simulation.
We generated photodisintegration events inside the sensitive volume of the MAIKo active target
and simulated various process such as ionization of the detection gas by decay particles, transport process of electrons, gas amplification on the GEM and $\mu$-PIC,
and response of the readout circuits.

First, the ionization process was simulated using the SRIM code~\cite{Ziegler2010}.
Second, the transport process of generated electrons was simulated with the Garfield++ code~\cite{Garfieldpp}
by using the drift velocity and diffusion coefficient of electrons calculated with the Magboltz code~\cite{Biagi1999}. 
Third, the gas amplification process of the electrons reached the GEM, and $\mu$-PIC was calculated 
with a simple stochastic model in which the amplification factor was assumed to be a random variable according to the P\'{o}lya distribution.  
Finally, the pseudo data set (hit pattern and signal shape) was virtually produced 
by a software that emulated the readout circuit~\cite{Takada2013}. 
The parameters for this simulation were optimized to reproduce the real data taken in the experiment.

The detection efficiency was estimated by analyzing the simulated pseudo data.
We defined the detection efficiency as the survival ratio of the produced events after the analysis.
We evaluated the efficiency at every polar and azimuthal angular bin with steps of $\Delta \theta = \Delta \phi = 20 ^{\circ}$ and
at energy bins with a step of $\Delta E_{\gamma} = 500 $~keV over $E_{\gamma} =$ 19--30~MeV.

\subsection{Beam Analysis}
\label{sec_beam}
The energy profile and total photon number of the beams irradiated on the MAIKo active target were estimated 
from the energy spectrum measured with the beam monitor located at the most downstream of the beam line.
A NaI(Tl) scintillation detector was employed as the beam monitor.

\begin{figure}[tbp]
    \includegraphics[width=86mm]{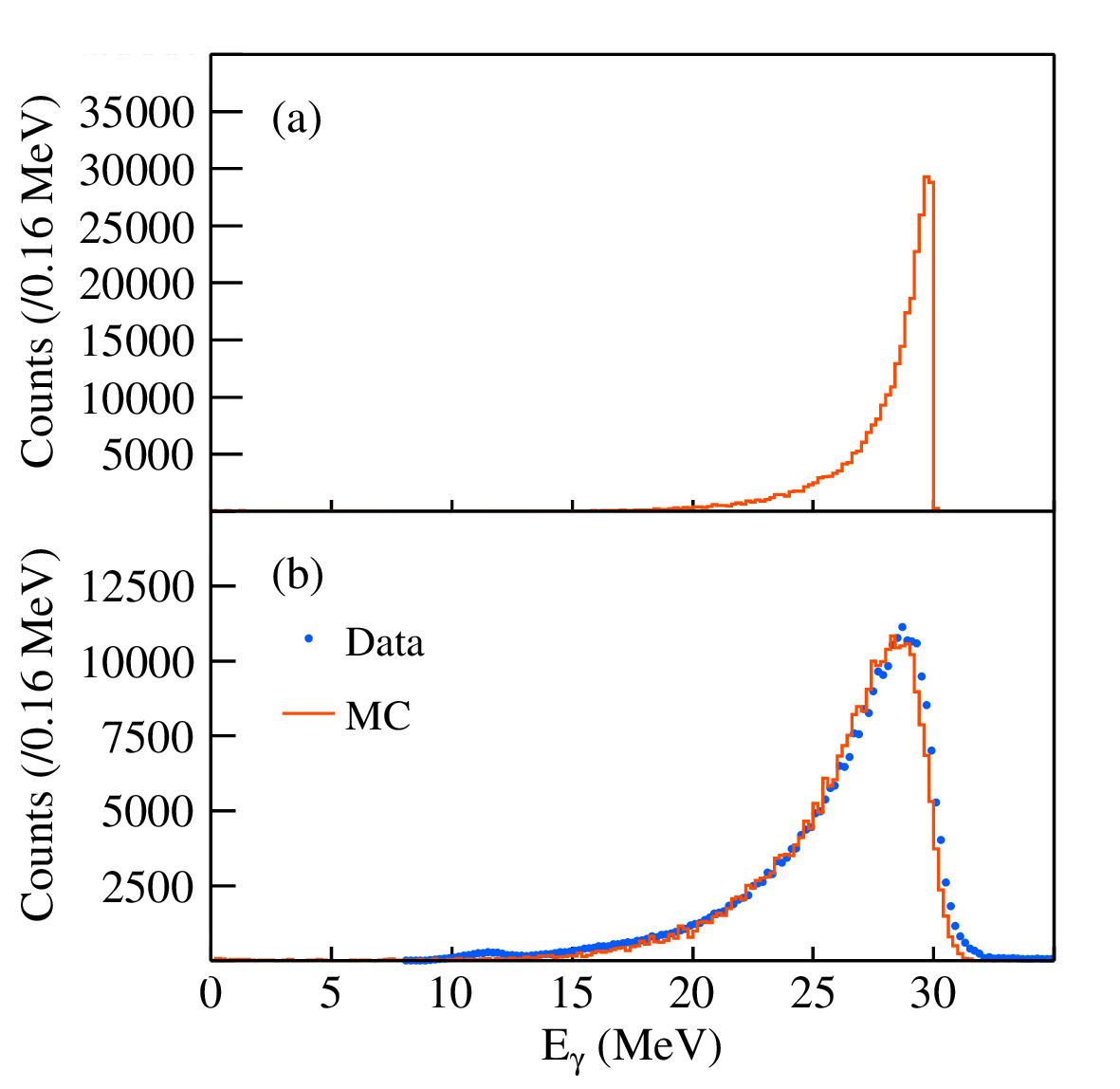}
    \caption{
    (a) Energy spectrum of the LCS photons estimated by the Monte Carlo simulation when the maximum LCS-photon energy is 30.0 MeV.
    (b) Simulated energy spectrum measured by the NaI(Tl) beam monitor compared with the experimental spectrum.
    The red solid line shows the simulated spectrum whereas the blue solid circles show the experimental spectrum.    
    }
    \label{fig_epro}
\end{figure}

The energy profile was estimated from a Monte Carlo simulation of the LCS beam generation process. 
We generated LCS photons at the collision point using realistic values of the parameters for the electron beam and the laser optical system, such as beam emittance and beam size.
We also virtually transported the LCS photons toward the NaI(Tl) beam monitor in the experimental hutch.
The interactions of the LCS photons with the beam-line materials were simulated by using the Geant4 toolkit~\cite{Agostinelli2003}.
The incident energy spectrum of the LCS photons estimated by the simulation is shown in Fig.~\ref{fig_epro}(a).
Its deduced energy-deposit spectrum folded with the finite energy resolution of the NaI(Tl) detector was
compared with the experimental spectrum measured when the beam intensity was as low as several thousands photons per second in Fig.~\ref{fig_epro}(b).

\begin{figure}[tbp]
    \includegraphics[width=86mm]{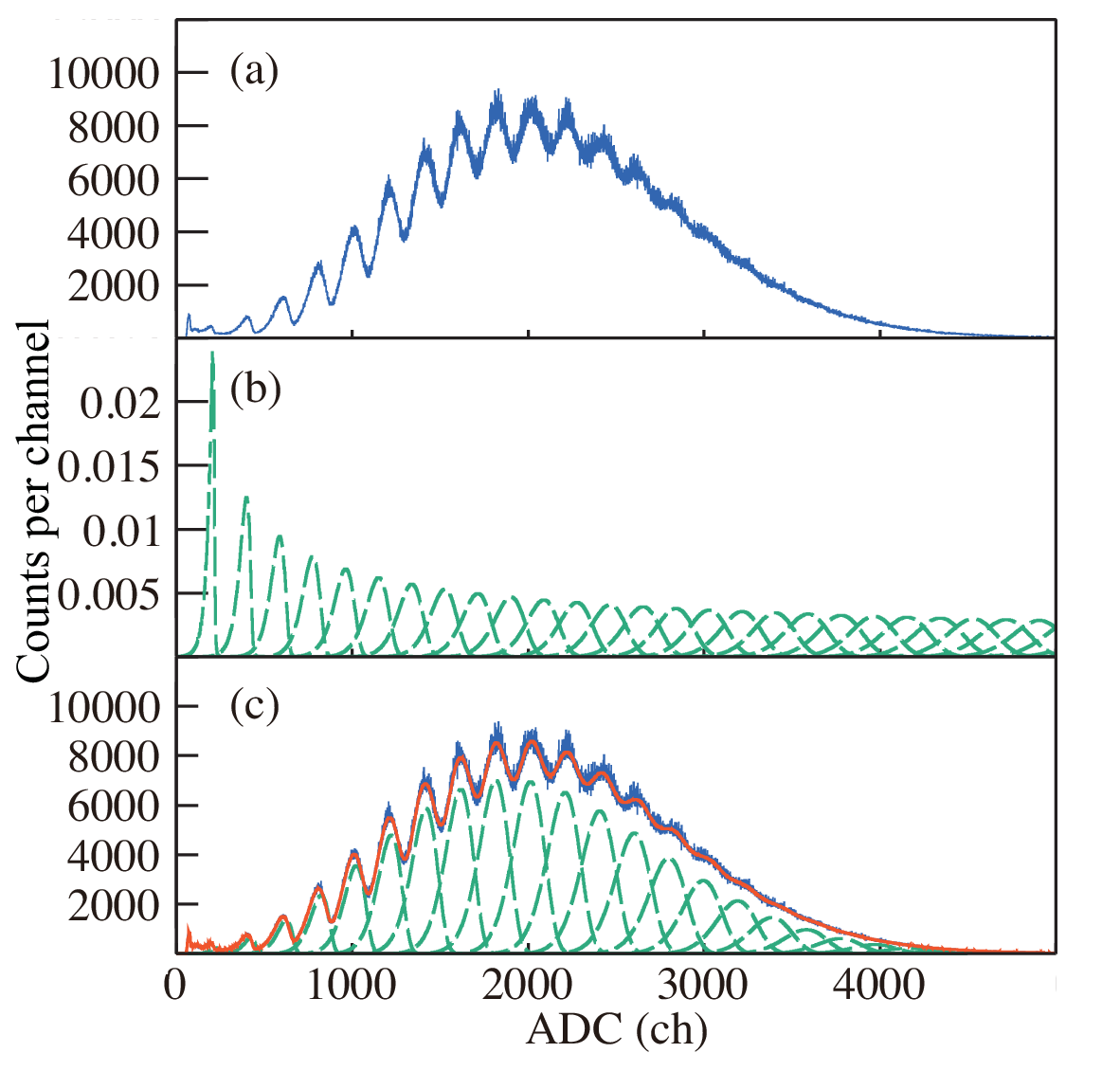}
    \caption{Typical procedure for estimating the LCS-photon number.
    (a) Energy spectrum measured with the NaI(Tl) beam monitor.
    (b) Response functions of the NaI(Tl) beam monitor for various photon multiplicities.
    (c) Best fit result (thick red solid line) to the measured spectrum (blue solid line) by the response functions multiplied by the weight factors (green dashed line).
    The quenching effect of the signals induced on the NaI(Tl) scintillator was taken into account (see text).
    }
    \label{fig_nphoton}
\end{figure}

Figure~\ref{fig_nphoton}(a) shows a typical energy spectrum measured with the beam monitor
when the maximum LCS-photon energy was 30.0 MeV, and the beam intensity was approximately 172~k photons per second.
Notably, several discrete peak structures were observed.
These structures were due to the multi-hit events in which multiple LCS photons were detected simultaneously.
The LCS photons intermittently arrived at the beam monitor at 16 kHz,
which was synchronized with the repetition frequency of the laser module for the LCS photon production.
In a single beam pulse, the number of photons varied between one and a few tens.
Because the time difference among the LCS photons in one pulse was shorter than the time resolution of the beam monitor,
these photons were detected as a single signal.
The discrete peaks in the spectrum correspond to one, two, three, and more photons in the order from left.
This distribution can be resolved as a linear combination of detector-response functions for various photon multiplicities.

The response function for one photon injection 
was obtained from the measurement with a low-intensity beam.
The multi-photon response templates of up to 60 photons were also generated from the iterative convolutions of the one photon response.
Figure~\ref{fig_nphoton}(b) shows the response functions for various photon multiplicities.
These functions are normalized so that the integral values are equal to unity.
Moreover, the response function of the background events caused by a source other than the LCS
was generated from the measurement with the laser turned off. 
The counts at the $x$ ch in the measured energy spectrum should be described by the template function $T(x)$ composed of a linear combination of the detector-response functions and the background response as follows
\begin{equation}
T(x) = \sum_{n=1}^{60}w_n t_n(x) + w_{\mathrm{BG}}t_{\mathrm{BG}}(x),
\end{equation}
where $t_{n}(x)$ and $t_{\mathrm{BG}}(x)$ stand for the $i$-photons and the background response functions,
and $w_i$ and $w_{\mathrm{BG}}$ are the weighting factors for the response functions.
In addition, the quenching effect of the signals induced on the NaI(Tl) scintillator was also taken into account
by distorting $T(x)$ using the empirical saturation function described in Ref.~\cite{Utsunomiya2018}.
We fitted this function to the measured energy spectra to determine the weighting factors and the parameters for the saturation function reproducing the spectra best. 
Figure~\ref{fig_nphoton}(c) shows the best-fit template functions compared with the measured spectrum.
The distorted response functions multiplied by the weight factors are also displayed with green dashed lines.

The total photon number irradiated on the MAIKo active target $N_{\gamma}$ was determined as 
\begin{equation}
N_{\gamma} = \sum_{n=1}^{60} n w_n,
\end{equation}
by using the weighting factors determined from the fitting.
The statistical uncertainty of the total photon number was estimated to be approximately 0.2\%.
The systematic uncertainty originated from the fitting procedure,
the quenching of detector output, the contribution from the background events, and the photon multiplicity higher than 60 was approximately 4\%.

\subsection{Cross Sections}
Energy averaged total cross sections of the photodisintegrations were determined from the yield ($Y$), the detection efficiency ($\epsilon$), and total photon number ($N_{\gamma}$).
Here, we define the energy-profile function $f_k(E)$ where $k$ is the index of the beam energy listed in Table~\ref{tab_measurement}.
The energy-profile function $f_{k}(E)$ was estimated by the Monte Carlo simulation described in Sec.~\ref{sec_beam} [see Fig.~\ref{fig_epro}(a)]
and normalized so that the energy-integral value was equal to the total photon number $N_{\gamma, k}$ as
\begin{equation}
    N_{\gamma, k} = \int_{0}^{E_{\mathrm{max}, k}} dE f_k(E),
\end{equation}
where $E_{\mathrm{max},k}$ denotes the maximum energy of the photon beam.

In preparation for determining the total cross sections, we first evaluated the differential cross section
at each angular bin using the following formula
\begin{equation}
\label{eq_diff_cs}
\left\langle \frac{d\sigma}{d\Omega} \right\rangle_{\theta_i, \phi_j, k} =  \frac{Y_{i, j, k}}
{\left[\int_{0}^{E_{\mathrm{max},k}}dE~f_{k}(E) \epsilon_{i, j, k}(E)  \right] ~ \tau \Delta \Omega_{i,j}},
\end{equation}
where $i$ and $j$ are the indices of the angular bins of $\theta$ and $\phi$ whereas $k$ is the index of the beam energy.
Here $\tau$ is the target thickness given in the areal number density of $^4$He nuclei,
and $\Delta \Omega_{i, j}$ is the solid angle of the angular bin 
with $\Delta \phi = \Delta \theta = 20^{\circ}$ around $\theta_i$ and $\phi_j$, respectively.

The total cross section is obtained by integrating the differential cross sections over the full solid angle of 4$\pi$.
In the present analysis, however, the cross sections in the angular bins with lower detection efficiencies must be removed from the integration
because those angular bins caused larger systematic uncertainties amplified by the efficiencies.
Hereafter, we define $\Delta\tilde{\Omega}_k$ as the angular range in which the differential cross sections are reliable and included in the analysis.

In order to obtain the total cross section, it is necessary to deduce the correction factor for converting the cross section
integrated over the partial solid angle $\Delta \tilde{\Omega}_k$ into that over $4\pi$. 
For this purpose, we assume the photodisintegration reactions in the present energy region are primarily induced through the $E1$ transition,
and model the angular distribution of the differential cross sections at $E_\gamma = E$ by the following formula
\begin{equation}
    \label{eq_ang_dep}
    \left(\frac{d\sigma_{M}}{d\Omega}\right)_{k} = \sigma_{M}(E) \left\{\frac{3}{8\pi}\sin ^2 \theta \left[ 1+\alpha_{k} \cos 2(\phi - \tilde{\phi)}\right] \right\},
\end{equation} 
where $\theta$ and $\phi$ are the polar and azimuthal angles of the decay particles in the center of mass frame.
$\tilde{\phi}$ and $\alpha_{k}$ are the polarization direction of the linearly polarized photon beam and the azimuthal asymmetry parameter, respectively.
$\sigma_M$ corresponds to the total cross section obtained by integrating the modeled differential cross section over $4\pi$.
In addition, we assume that the energy dependence of the modeled total cross section $\sigma_M(E)$ is given by 
\begin{equation}
\label{eq_e_dep}
\sigma_{M} (E) = 
\begin{dcases}
    \sum_{l=1}^{4}p_l(E-E_{\mathrm{th}})^l  & (E \geq E_{\mathrm{th}}) \\
    0 & (E<E_{\mathrm{th}}),
\end{dcases}
\end{equation}
where $E_{\mathrm{th}}$ is the threshold energy of the photodisintegration reactions.

Based on the cross section model described above, the expected yield $\tilde{Y}$ at a certain angular bin is estimated as
\begin{equation}
    \label{eq_yield_est}
\tilde{Y}_{\theta_i, \phi_j, k} = \tau \int d\Omega \int_{0}^{E_{\mathrm{max},k}} dE \left[ I_{\Delta\Omega_{i, j}}\left( \frac{d\sigma_{M}}{d\Omega}\right)_k  f_{k} \epsilon_{i, j, k} \right].
\end{equation}
Here, for the sake of simplicity, the arguments ($E$, $\theta$, and $\phi$) of the functions in the integrand are omitted.
$I_A$ is an indicator function on the angular range which is defined as,
\begin{equation}
    I_{A}(x) = 
    \begin{cases}
        1 & (x \in A)\\
        0 & (x \notin A).
    \end{cases}
\end{equation}
Therefore, the angular integration in Eq.~(\ref{eq_yield_est}) was done within an angular bin $\Delta \Omega_{i, j}$
with $\Delta\phi=\Delta\theta=20^\circ$ around $\theta_i$ and $\phi_j$.
This angular bin is the same as that employed in Eq.~(\ref{eq_diff_cs}).
The unknown parameters, $\alpha_k$ and $p_l$ in Eqs.~(\ref{eq_ang_dep}) and (\ref{eq_e_dep}), were optimized to maximize the likelihood of the expected yields $\tilde{Y}$
by comparing them with the measured yields $Y$ at all angular bins and beam energies while $\tilde{\phi}$ was fixed at $\tilde{\phi} = 10 ^\circ$.

The average beam energy $\langle E \rangle_k$ was calculated
from the weighted mean of the beam energy defined as follows:
\begin{equation}
    \label{eq_ave_e}
    \langle E \rangle_{k} = \frac{\int_{0}^{E_{\mathrm{max}, k}} dE \int d\Omega \left\{ E \left[ I_{\Delta \tilde{\Omega}_k}   \left(\frac{d\sigma_{M}}{d\Omega}\right)_{k} f_k  \epsilon_{i, j, k} \right]\right\}}
    {\int_{0}^{E_{\mathrm{max}, k}} dE \int d\Omega \left[ I_{\Delta \tilde{\Omega}_k} \left(\frac{d\sigma_M}{d\Omega}\right)_{k} f_k  \epsilon_{i, j, k} \right]}.
\end{equation}

Finally, the measured cross sections were corrected to obtain the experimental total cross section at the average beam energy $\langle E \rangle_k$ by the following formula
\begin{eqnarray}
    \label{eq_tot_cs}
\langle \sigma \rangle _{k} &=& C_k \left\langle \sigma\right\rangle_{\Delta\tilde{\Omega}_k}\\
                            &=& C_k \sum^{\Delta \tilde{\Omega}_k}_{\theta_i, \phi_j} \left\langle \frac{d\sigma}{d\Omega} \right\rangle_{\theta_i, \phi_j,k} \Delta\Omega_{i,j}
\end{eqnarray}
where the correction factor $C_k$ is given as 
\begin{equation}
C_k = \frac{\sigma_M(\langle E \rangle_k)}{\int d\Omega \left[ I_{\Delta \tilde{\Omega}_k} \left( \frac{d\sigma_{M} }{d\Omega}\right)_k (\langle E \rangle_k) \right]}.
\end{equation}
The second factor in Eq.~(\ref{eq_tot_cs}), $ \left\langle \sigma\right\rangle_{\Delta\tilde{\Omega}_k}$ is the integration of the differential cross section in Eq.~(\ref{eq_diff_cs}) over the angular bins adopted in the analysis.
The correction factor $C_k$ corresponds to the ratio of the modeled cross section integrated over $4\pi$ to that over $\Delta \tilde{\Omega}_k$.
The uncertainty of the total cross sections due to the assumption on the angular distribution of the differential cross section was evaluated 
by using a general formula given in Ref.~\cite{Shima2005} instead of Eq.(\ref{eq_ang_dep}),
and that due to the definition of $\Delta \tilde{\Omega}_k$ was also estimated 
by varying $\Delta \tilde{\Omega}_k$.
They were included in the systematic uncertainty of the total cross sections.

\section{Results \& Discussion}

\begin{table*}
\caption{\label{tab_cs} Experimental total cross sections for the $^4$He($\gamma,n$)$^3$He and $^4$He($\gamma, p$)$^3$H reactions.
The average beam energies $\langle E \rangle_k$ obtained from Eq. (\ref{eq_ave_e}) and  
the statistical and systematic uncertainties are also presented.
$k$ is an index of beam energy defined in Table~\ref{tab_measurement}.}
\begin{ruledtabular}
\begin{tabular}{l c c c c c c c c c}
    && \multicolumn{4}{c}{$^4$He$(\gamma, n)$$^3$He}& \multicolumn{4}{c}{$^4$He$(\gamma, p)$$^3$H}\\ \cline{3-6} \cline{7-10}
    $k$ & $E_{\mathrm{max}}$ & $\left\langle E \right\rangle_k$ &  $\langle \sigma \rangle_k$ & $\Delta \sigma_{\mathrm{stat}}$  & $\Delta \sigma _{\mathrm{sys}}$  & 
    $\left\langle E \right\rangle_k$ &  $\langle \sigma \rangle_k$ & $\Delta \sigma_{\mathrm{stat}}$ & $\Delta \sigma _{\mathrm{sys}}$ \\
     & (MeV) & (MeV) &   (mb) & (mb) &  (mb) & (MeV) &   (mb) &  (mb) &  (mb)\\
    \hline
    1 & 23.0 & 22.5 & 1.15 & $\pm$ 0.08 &  $+ 0.13 - 0.13$ & 22.5 & 1.25 & $\pm$ 0.08 &  $+ 0.56 - 0.40$ \\ 
    2 & 24.0 & 23.5 & 1.31 & $\pm$ 0.10 &  $+ 0.14 - 0.15$ & 23.3 & 1.41 & $\pm$ 0.07 &  $+ 0.28 - 0.28$ \\ 
    3 & 25.0 & 24.5 & 1.66 & $\pm$ 0.12 &  $+ 0.18 - 0.18$ & 24.3 & 1.64 & $\pm$ 0.08 &  $+ 0.25 - 0.22$ \\ 
    4 & 27.0 & 26.5 & 2.26 & $\pm$ 0.25 &  $+ 0.25 - 0.24$ & 25.7 & 1.93 & $\pm$ 0.13 &  $+ 0.34 - 0.42$ \\ 
    5 & 28.0 & 27.5 & 1.94 & $\pm$ 0.19 &  $+ 0.21 - 0.24$ & 26.5 & 1.59 & $\pm$ 0.09 &  $+ 0.63 - 0.21$ \\ 
    6 & 30.0 & 29.4 & 1.26 & $\pm$ 0.11 &  $+ 0.16 - 0.19$ & 28.7 & 1.67 & $\pm$ 0.05 &  $+ 0.19 - 0.23$
\end{tabular}
\end{ruledtabular}
\end{table*}

\begin{figure}[tbp]
    \includegraphics[width=86mm]{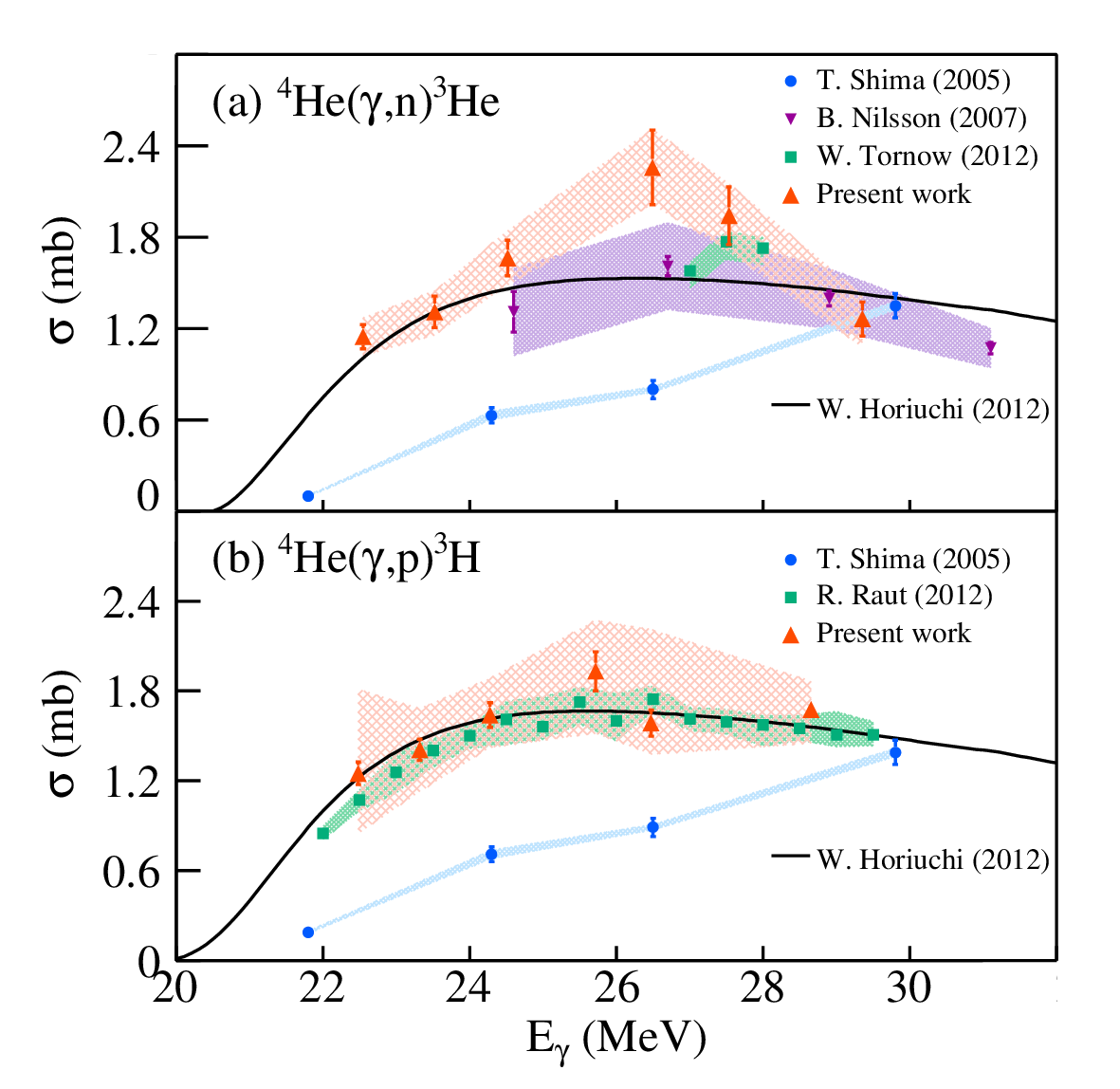}
    \caption{
        Experimental cross sections of the (a) $^4$He($\gamma,n$)$^3$He and (b) $^4$He($\gamma, p$)$^3$H reactions.
        The present result (upward triangles) is compared with those from previous studies (\cite{Shima2005}: circles,~\cite{Nilsson2007}: downward triangles,~\cite{Tornow2012} and \cite{Raut2012}: squares). 
        Systematic uncertainties of the cross sections are indicated with hatched regions.
        The theoretical predictions taken from Ref.~\cite{Horiuchi2012} are shown by the solid lines.
    }
    \label{fig_cross_sections}
\end{figure}

The total cross section of the $^4$He($\gamma, n$)$^3$He and $^4$He($\gamma, p$)$^3$H reactions measured in the present work are given in Table \ref{tab_cs} and plotted using the upward triangles in Fig.~\ref{fig_cross_sections}.
The statistical and systematic uncertainties are shown as the bars and the bands, respectively.

The systematic uncertainties considered in the present study are summarized below.
The uncertainty, which is primarily due to the correction of the solid angle and the model selection, is estimated to be varying between ten and a few tens of percentages, depending on the measurement conditions.
In addition, the uncertainty from the efficiency estimations due to the incompleteness of the Monte Carlo simulations was approximately 10\%, which is taken from Ref.~\cite{Furuno2019}.
As discussed in Sec.~\ref{sec_beam}, the uncertainty of the incident photon number was 4\%.
We monitored and controlled the density of the target gas, and uncertainty originated from that was 0.1\% or less.

The experimental total cross sections measured with the quasi-mono-energetic photon beams~\cite{Shima2005, Nilsson2007, Raut2012, Tornow2012} were selected
from numerous previous studies and compared with the present results in Fig. \ref{fig_cross_sections}. 
The present results showing the GDR peak structure around $E_{\gamma}=26$~MeV in both the $^4$He($\gamma, n$)$^3$He and $^4$He($\gamma, p$)$^3$H reactions
do not support those reported by Shima {\it et al.}~\cite{Shima2005}, 
but are consistent with those from the HI$\gamma$S group~\cite{Raut2012,Tornow2012}.
The uncertainties in the present results are slightly higher than those in previous studies,
but smaller than the difference between the results by Shima {\it et al.} and those by the Hi$\gamma$S group. 
Notably, the present $^4$He$(\gamma,n)$$^3$He result also agrees
with that by the MAX-lab group~\cite{Nilsson2007},
which was obtained by using energy-tagged bremsstrahlung photons.
The GDR peak energy is in accord with the theoretical calculation~\cite{Horiuchi2012} shown by the solid lines in Fig~\ref{fig_cross_sections}. 
In addition, the absolute values of our data support the theoretical calculations within the uncertainty
except for some points of the $^4$He($\gamma, n$)$^3$He reaction whereby their statistical uncertainties are higher than those of others.

As described in Sec.~\ref{sec_intro}, the cross sections of the $^4$He photodisintegration reactions in the GDR energy region are
important to elucidate the certain processes in the nucleosynthesis in the universe.
Because cross sections theoretically calculated from the nuclear structure theory are generally employed for the astrophysical calculation,
one would have to revise the theoretical frameworks and the scenario of the nucleosynthesis
if the theoretical predictions largely deviate from the experimental cross sections.
The striking experimental results reported by Shima {\it et al.} challenged the conventional view of 
the $^4$He photodisintegration reactions that the GDR peak was located around $E_{\gamma}=26$ MeV.
However, the present results support the conventional view and 
do not force one to revise the physical pictures on the nucleosynthesis drawn by the nuclear structure theories.

\section{Summary}

We performed a simultaneous measurement of the $^4$He($\gamma$, $n$)$^3$He and $^4$He($\gamma, p$)$^3$H 
reactions at the BL01 in the NewSUBARU synchrotron radiation facility
by using the TPC-based gaseous active target, MAIKo.
Quasi-mono-energetic photon beams at $E_{\gamma} = $23--30~MeV, 
which were close to the GDR energy in $^4$He,
were irradiated on the MAIKo active target.
Three-dimensional trajectories of charged particles emitted from the photonuclear reactions were measured, 
and the $^4$He photodisintegration events were identified.
The detection efficiencies of the photodisintegration events by the MAIKo active target were estimated with the Monte Carlo simulation
in which the response of the detector was considered. 
The total photon number and energy distribution of photon beams were evaluated from 
the energy spectra measured with the NaI(Tl) scintillator.
Finally, we determined the cross sections of the $^4$He($\gamma, n$)$^3$He and $^4$He($\gamma, p$)$^3$H reactions.

The cross sections obtained from the present measurement show the GDR peaks around $E_{\gamma} = $ 26~MeV.
This result is in accord with the experimental results from the HI$\gamma$S facility~\cite{Tornow2012, Raut2012} and the MAX-lab facility~\cite{Nilsson2007}
but inconsistent with the result from Shima {\it et al.}~\cite{Shima2005}. 
The present results support the theoretical predictions for the $^4$He photodisintegration reactions~\cite{Quaglioni2004, Horiuchi2012},
and do not force one to revise the theoretical frameworks.

This study demonstrated the applicability of the MAIKo active target,
which was originally designed for the in-beam spectroscopy of the unstable nuclei~\cite{Furuno2018},
to the study of the photonuclear reactions.
The present work expanded our capability of providing reliable data on important reactions in nuclear astrophysics
with the MAIKo active target.

\begin{acknowledgments}
We express our gratitude to the NewSUBARU synchrotron radiation facility 
and Research Center for Nuclear Physics, Osaka University, 
for providing their facilities and supports for the present work.
We would like to thank  Prof. T.~Kajino, Prof. T.~Suzuki, and Prof. W. Horiuchi 
for fruitful discussions regarding interpretation of the experimental results.
M. M. appreciates the support of Grant-in-Aid for JSPS Research Fellow JP16J05592.
This work was also supported by JSPS KAKENHI Grants Numbers JP23340068 and JP15H02091.
\end{acknowledgments}

\bibliography{main}

\begin{thebibliography}{52}%
\makeatletter
\providecommand \@ifxundefined [1]{%
 \@ifx{#1\undefined}
}%
\providecommand \@ifnum [1]{%
 \ifnum #1\expandafter \@firstoftwo
 \else \expandafter \@secondoftwo
 \fi
}%
\providecommand \@ifx [1]{%
 \ifx #1\expandafter \@firstoftwo
 \else \expandafter \@secondoftwo
 \fi
}%
\providecommand \natexlab [1]{#1}%
\providecommand \enquote  [1]{``#1''}%
\providecommand \bibnamefont  [1]{#1}%
\providecommand \bibfnamefont [1]{#1}%
\providecommand \citenamefont [1]{#1}%
\providecommand \href@noop [0]{\@secondoftwo}%
\providecommand \href [0]{\begingroup \@sanitize@url \@href}%
\providecommand \@href[1]{\@@startlink{#1}\@@href}%
\providecommand \@@href[1]{\endgroup#1\@@endlink}%
\providecommand \@sanitize@url [0]{\catcode `\\12\catcode `\$12\catcode
  `\&12\catcode `\#12\catcode `\^12\catcode `\_12\catcode `\%12\relax}%
\providecommand \@@startlink[1]{}%
\providecommand \@@endlink[0]{}%
\providecommand \url  [0]{\begingroup\@sanitize@url \@url }%
\providecommand \@url [1]{\endgroup\@href {#1}{\urlprefix }}%
\providecommand \urlprefix  [0]{URL }%
\providecommand \Eprint [0]{\href }%
\providecommand \doibase [0]{https://doi.org/}%
\providecommand \selectlanguage [0]{\@gobble}%
\providecommand \bibinfo  [0]{\@secondoftwo}%
\providecommand \bibfield  [0]{\@secondoftwo}%
\providecommand \translation [1]{[#1]}%
\providecommand \BibitemOpen [0]{}%
\providecommand \bibitemStop [0]{}%
\providecommand \bibitemNoStop [0]{.\EOS\space}%
\providecommand \EOS [0]{\spacefactor3000\relax}%
\providecommand \BibitemShut  [1]{\csname bibitem#1\endcsname}%
\let\auto@bib@innerbib\@empty
\bibitem [{\citenamefont {Bohr}\ and\ \citenamefont
  {Mottelson}(1975)}]{Bohr1969}%
  \BibitemOpen
  \bibfield  {author} {\bibinfo {author} {\bibfnamefont {A.}~\bibnamefont
  {Bohr}}\ and\ \bibinfo {author} {\bibfnamefont {B.}~\bibnamefont
  {Mottelson}},\ }\href@noop {} {\emph {\bibinfo {title} {Nuclear Structure
  II}}}\ (\bibinfo  {publisher} {Benjamin Inc.},\ \bibinfo {year}
  {1975})\BibitemShut {NoStop}%
\bibitem [{\citenamefont {{Woosley}}\ \emph {et~al.}(1990)\citenamefont
  {{Woosley}}, \citenamefont {{Hartmann}}, \citenamefont {{Hoffman}},\ and\
  \citenamefont {{Haxton}}}]{Woosley1990}%
  \BibitemOpen
  \bibfield  {author} {\bibinfo {author} {\bibfnamefont {S.~E.}\ \bibnamefont
  {{Woosley}}}, \bibinfo {author} {\bibfnamefont {D.~H.}\ \bibnamefont
  {{Hartmann}}}, \bibinfo {author} {\bibfnamefont {R.~D.}\ \bibnamefont
  {{Hoffman}}},\ and\ \bibinfo {author} {\bibfnamefont {W.~C.}\ \bibnamefont
  {{Haxton}}},\ }\bibfield  {title} {\bibinfo {title} {{The $\nu$-Process}},\
  }\href {https://doi.org/10.1086/168839} {\bibfield  {journal} {\bibinfo
  {journal} {The Astrophysical Journal}\ }\textbf {\bibinfo {volume} {356}},\
  \bibinfo {pages} {272} (\bibinfo {year} {1990})}\BibitemShut {NoStop}%
\bibitem [{\citenamefont {Suzuki}\ and\ \citenamefont
  {Kajino}(2013)}]{Suzuki2013}%
  \BibitemOpen
  \bibfield  {author} {\bibinfo {author} {\bibfnamefont {T.}~\bibnamefont
  {Suzuki}}\ and\ \bibinfo {author} {\bibfnamefont {T.}~\bibnamefont
  {Kajino}},\ }\bibfield  {title} {\bibinfo {title} {Element synthesis in the
  supernova environment and neutrino oscillations},\ }\href
  {https://doi.org/10.1088/0954-3899/40/8/083101} {\bibfield  {journal}
  {\bibinfo  {journal} {Journal of Physics G: Nuclear and Particle Physics}\
  }\textbf {\bibinfo {volume} {40}},\ \bibinfo {pages} {083101} (\bibinfo
  {year} {2013})}\BibitemShut {NoStop}%
\bibitem [{\citenamefont {Nakamura}\ \emph {et~al.}(2017)\citenamefont
  {Nakamura}, \citenamefont {Kamano}, \citenamefont {Hayato}, \citenamefont
  {Hirai}, \citenamefont {Horiuchi}, \citenamefont {Kumano}, \citenamefont
  {Murata}, \citenamefont {Saito}, \citenamefont {Sakuda}, \citenamefont
  {Sato},\ and\ \citenamefont {Suzuki}}]{Nakamura2017}%
  \BibitemOpen
  \bibfield  {author} {\bibinfo {author} {\bibfnamefont {S.~X.}\ \bibnamefont
  {Nakamura}}, \bibinfo {author} {\bibfnamefont {H.}~\bibnamefont {Kamano}},
  \bibinfo {author} {\bibfnamefont {Y.}~\bibnamefont {Hayato}}, \bibinfo
  {author} {\bibfnamefont {M.}~\bibnamefont {Hirai}}, \bibinfo {author}
  {\bibfnamefont {W.}~\bibnamefont {Horiuchi}}, \bibinfo {author}
  {\bibfnamefont {S.}~\bibnamefont {Kumano}}, \bibinfo {author} {\bibfnamefont
  {T.}~\bibnamefont {Murata}}, \bibinfo {author} {\bibfnamefont
  {K.}~\bibnamefont {Saito}}, \bibinfo {author} {\bibfnamefont
  {M.}~\bibnamefont {Sakuda}}, \bibinfo {author} {\bibfnamefont
  {T.}~\bibnamefont {Sato}},\ and\ \bibinfo {author} {\bibfnamefont
  {Y.}~\bibnamefont {Suzuki}},\ }\bibfield  {title} {\bibinfo {title} {Towards
  a unified model of neutrino-nucleus reactions for neutrino oscillation
  experiments},\ }\href {https://doi.org/10.1088/1361-6633/aa5e6c} {\bibfield
  {journal} {\bibinfo  {journal} {Reports on Progress in Physics}\ }\textbf
  {\bibinfo {volume} {80}},\ \bibinfo {pages} {056301} (\bibinfo {year}
  {2017})}\BibitemShut {NoStop}%
\bibitem [{\citenamefont {Ejiri}\ \emph {et~al.}(2013)\citenamefont {Ejiri},
  \citenamefont {Titov}, \citenamefont {Boswell},\ and\ \citenamefont
  {Young}}]{Ejiri2013}%
  \BibitemOpen
  \bibfield  {author} {\bibinfo {author} {\bibfnamefont {H.}~\bibnamefont
  {Ejiri}}, \bibinfo {author} {\bibfnamefont {A.~I.}\ \bibnamefont {Titov}},
  \bibinfo {author} {\bibfnamefont {M.}~\bibnamefont {Boswell}},\ and\ \bibinfo
  {author} {\bibfnamefont {A.}~\bibnamefont {Young}},\ }\bibfield  {title}
  {\bibinfo {title} {Neutrino-nuclear response and photonuclear reactions},\
  }\href {https://doi.org/10.1103/PhysRevC.88.054610} {\bibfield  {journal}
  {\bibinfo  {journal} {Phys. Rev. C}\ }\textbf {\bibinfo {volume} {88}},\
  \bibinfo {pages} {054610} (\bibinfo {year} {2013})}\BibitemShut {NoStop}%
\bibitem [{\citenamefont {Kusakabe}\ \emph {et~al.}(2009)\citenamefont
  {Kusakabe}, \citenamefont {Kajino}, \citenamefont {Yoshida}, \citenamefont
  {Shima}, \citenamefont {Nagai},\ and\ \citenamefont {Kii}}]{Kusakabe2008}%
  \BibitemOpen
  \bibfield  {author} {\bibinfo {author} {\bibfnamefont {M.}~\bibnamefont
  {Kusakabe}}, \bibinfo {author} {\bibfnamefont {T.}~\bibnamefont {Kajino}},
  \bibinfo {author} {\bibfnamefont {T.}~\bibnamefont {Yoshida}}, \bibinfo
  {author} {\bibfnamefont {T.}~\bibnamefont {Shima}}, \bibinfo {author}
  {\bibfnamefont {Y.}~\bibnamefont {Nagai}},\ and\ \bibinfo {author}
  {\bibfnamefont {T.}~\bibnamefont {Kii}},\ }\bibfield  {title} {\bibinfo
  {title} {New constraints on radiative decay of long-lived particles in big
  bang nucleosynthesis with new $^{4}\mathrm{He}$ photodisintegration data},\
  }\href {https://doi.org/10.1103/PhysRevD.79.123513} {\bibfield  {journal}
  {\bibinfo  {journal} {Phys. Rev. D}\ }\textbf {\bibinfo {volume} {79}},\
  \bibinfo {pages} {123513} (\bibinfo {year} {2009})}\BibitemShut {NoStop}%
\bibitem [{\citenamefont {Ryan}\ \emph {et~al.}(2000)\citenamefont {Ryan},
  \citenamefont {Beers}, \citenamefont {Olive}, \citenamefont {Fields},\ and\
  \citenamefont {Norris}}]{Ryan2000}%
  \BibitemOpen
  \bibfield  {author} {\bibinfo {author} {\bibfnamefont {S.~G.}\ \bibnamefont
  {Ryan}}, \bibinfo {author} {\bibfnamefont {T.~C.}\ \bibnamefont {Beers}},
  \bibinfo {author} {\bibfnamefont {K.~A.}\ \bibnamefont {Olive}}, \bibinfo
  {author} {\bibfnamefont {B.~D.}\ \bibnamefont {Fields}},\ and\ \bibinfo
  {author} {\bibfnamefont {J.~E.}\ \bibnamefont {Norris}},\ }\bibfield  {title}
  {\bibinfo {title} {Primordial lithium and big bang nucleosynthesis},\ }\href
  {https://doi.org/10.1086/312492} {\bibfield  {journal} {\bibinfo  {journal}
  {The Astrophysical Journal}\ }\textbf {\bibinfo {volume} {530}},\ \bibinfo
  {pages} {L57} (\bibinfo {year} {2000})}\BibitemShut {NoStop}%
\bibitem [{\citenamefont {Mel^^c3^^a9ndez}\ and\ \citenamefont
  {Ram^^c3^^adrez}(2004)}]{Melendez2004}%
  \BibitemOpen
  \bibfield  {author} {\bibinfo {author} {\bibfnamefont {J.}~\bibnamefont
  {Mel^^c3^^a9ndez}}\ and\ \bibinfo {author} {\bibfnamefont {I.}~\bibnamefont
  {Ram^^c3^^adrez}},\ }\bibfield  {title} {\bibinfo {title} {Reappraising the
  spite lithium plateau: Extremely thin and marginally consistent with {WMAP}
  data},\ }\href {https://doi.org/10.1086/425962} {\bibfield  {journal}
  {\bibinfo  {journal} {The Astrophysical Journal}\ }\textbf {\bibinfo {volume}
  {615}},\ \bibinfo {pages} {L33} (\bibinfo {year} {2004})}\BibitemShut
  {NoStop}%
\bibitem [{\citenamefont {Asplund}\ \emph {et~al.}(2006)\citenamefont
  {Asplund}, \citenamefont {Lambert}, \citenamefont {Nissen}, \citenamefont
  {Primas},\ and\ \citenamefont {Smith}}]{Asplund2006}%
  \BibitemOpen
  \bibfield  {author} {\bibinfo {author} {\bibfnamefont {M.}~\bibnamefont
  {Asplund}}, \bibinfo {author} {\bibfnamefont {D.~L.}\ \bibnamefont
  {Lambert}}, \bibinfo {author} {\bibfnamefont {P.~E.}\ \bibnamefont {Nissen}},
  \bibinfo {author} {\bibfnamefont {F.}~\bibnamefont {Primas}},\ and\ \bibinfo
  {author} {\bibfnamefont {V.~V.}\ \bibnamefont {Smith}},\ }\bibfield  {title}
  {\bibinfo {title} {Lithium isotopic abundances in metal-poor halo stars},\
  }\href {https://doi.org/10.1086/503538} {\bibfield  {journal} {\bibinfo
  {journal} {The Astrophysical Journal}\ }\textbf {\bibinfo {volume} {644}},\
  \bibinfo {pages} {229} (\bibinfo {year} {2006})}\BibitemShut {NoStop}%
\bibitem [{\citenamefont {Gorbunov}\ and\ \citenamefont
  {Spiridonov}(1957)}]{Gorbunov1958_1}%
  \BibitemOpen
  \bibfield  {author} {\bibinfo {author} {\bibfnamefont {A.~N.}\ \bibnamefont
  {Gorbunov}}\ and\ \bibinfo {author} {\bibfnamefont {V.~M.}\ \bibnamefont
  {Spiridonov}},\ }\bibfield  {title} {\bibinfo {title} {Photodisintegration of
  helium, i.},\ }\href@noop {} {\bibfield  {journal} {\bibinfo  {journal} {Zh.
  Eksp. Teor. Fiz.}\ }\textbf {\bibinfo {volume} {33}},\ \bibinfo {pages} {21}
  (\bibinfo {year} {1957})},\ \bibinfo {note} {[Sov. Phys. JETP \textbf{6}, 16
  (1958)]}\BibitemShut {NoStop}%
\bibitem [{\citenamefont {Gorbunov}\ and\ \citenamefont
  {Spiridonov}(1958)}]{Gorbunov1958_2}%
  \BibitemOpen
  \bibfield  {author} {\bibinfo {author} {\bibfnamefont {A.~N.}\ \bibnamefont
  {Gorbunov}}\ and\ \bibinfo {author} {\bibfnamefont {V.~M.}\ \bibnamefont
  {Spiridonov}},\ }\bibfield  {title} {\bibinfo {title} {Photodisintegration of
  helium. ii},\ }\href@noop {} {\bibfield  {journal} {\bibinfo  {journal} {Zh.
  Eksp. Teor. Fiz.}\ }\textbf {\bibinfo {volume} {34}},\ \bibinfo {pages} {862}
  (\bibinfo {year} {1958})},\ \bibinfo {note} {[Sov. Phys. JETP \textbf{7}, 596
  (1958)]}\BibitemShut {NoStop}%
\bibitem [{\citenamefont {Gardner}\ and\ \citenamefont
  {Anderson}(1962)}]{Gardner1962}%
  \BibitemOpen
  \bibfield  {author} {\bibinfo {author} {\bibfnamefont {C.~C.}\ \bibnamefont
  {Gardner}}\ and\ \bibinfo {author} {\bibfnamefont {J.~D.}\ \bibnamefont
  {Anderson}},\ }\bibfield  {title} {\bibinfo {title} {Gamma yield from the
  proton bombardment of tritium},\ }\href
  {https://doi.org/10.1103/PhysRev.125.626} {\bibfield  {journal} {\bibinfo
  {journal} {Phys. Rev.}\ }\textbf {\bibinfo {volume} {125}},\ \bibinfo {pages}
  {626} (\bibinfo {year} {1962})}\BibitemShut {NoStop}%
\bibitem [{\citenamefont {Gemmell}\ and\ \citenamefont
  {Jones}(1962)}]{Gemmell1962}%
  \BibitemOpen
  \bibfield  {author} {\bibinfo {author} {\bibfnamefont {D.}~\bibnamefont
  {Gemmell}}\ and\ \bibinfo {author} {\bibfnamefont {G.}~\bibnamefont
  {Jones}},\ }\bibfield  {title} {\bibinfo {title} {The {T($p,\gamma$)He$^4$}
  reaction},\ }\href
  {https://doi.org/https://doi.org/10.1016/0029-5582(62)90508-4} {\bibfield
  {journal} {\bibinfo  {journal} {Nuclear Physics}\ }\textbf {\bibinfo {volume}
  {33}},\ \bibinfo {pages} {102} (\bibinfo {year} {1962})}\BibitemShut
  {NoStop}%
\bibitem [{\citenamefont {Clerc}\ \emph {et~al.}(1965)\citenamefont {Clerc},
  \citenamefont {Stewart},\ and\ \citenamefont {Morrison}}]{Clerc1965}%
  \BibitemOpen
  \bibfield  {author} {\bibinfo {author} {\bibfnamefont {H.}~\bibnamefont
  {Clerc}}, \bibinfo {author} {\bibfnamefont {R.}~\bibnamefont {Stewart}},\
  and\ \bibinfo {author} {\bibfnamefont {R.}~\bibnamefont {Morrison}},\
  }\bibfield  {title} {\bibinfo {title} {Photodisintegration of {$^4$He}},\
  }\href {https://doi.org/https://doi.org/10.1016/0031-9163(65)90353-7}
  {\bibfield  {journal} {\bibinfo  {journal} {Physics Letters}\ }\textbf
  {\bibinfo {volume} {18}},\ \bibinfo {pages} {316} (\bibinfo {year}
  {1965})}\BibitemShut {NoStop}%
\bibitem [{\citenamefont {Gorbunov}(1968)}]{Gorbunov1968}%
  \BibitemOpen
  \bibfield  {author} {\bibinfo {author} {\bibfnamefont {A.}~\bibnamefont
  {Gorbunov}},\ }\bibfield  {title} {\bibinfo {title} {Study of the {$^4$He
  ($\gamma, p$)$^3$H and $^4$He($\gamma, n$)$^3$He} reactions},\ }\href
  {https://doi.org/https://doi.org/10.1016/0370-2693(68)90230-X} {\bibfield
  {journal} {\bibinfo  {journal} {Physics Letters B}\ }\textbf {\bibinfo
  {volume} {27}},\ \bibinfo {pages} {436} (\bibinfo {year} {1968})}\BibitemShut
  {NoStop}%
\bibitem [{\citenamefont {Meyerhof}\ \emph {et~al.}(1970)\citenamefont
  {Meyerhof}, \citenamefont {Suffert},\ and\ \citenamefont
  {Feldman}}]{Meyerhof1970}%
  \BibitemOpen
  \bibfield  {author} {\bibinfo {author} {\bibfnamefont {W.~E.}\ \bibnamefont
  {Meyerhof}}, \bibinfo {author} {\bibfnamefont {M.}~\bibnamefont {Suffert}},\
  and\ \bibinfo {author} {\bibfnamefont {W.}~\bibnamefont {Feldman}},\
  }\bibfield  {title} {\bibinfo {title} {{$^3$H($p,\gamma$)$^4$He} reaction
  from 3 to 18 {MeV}},\ }\href
  {https://doi.org/https://doi.org/10.1016/0375-9474(70)90619-6} {\bibfield
  {journal} {\bibinfo  {journal} {Nuclear Physics A}\ }\textbf {\bibinfo
  {volume} {148}},\ \bibinfo {pages} {211} (\bibinfo {year}
  {1970})}\BibitemShut {NoStop}%
\bibitem [{\citenamefont {Sanada}\ \emph {et~al.}(1970)\citenamefont {Sanada},
  \citenamefont {Yamanouchi}, \citenamefont {Sakai},\ and\ \citenamefont
  {Seki}}]{Sanada1970}%
  \BibitemOpen
  \bibfield  {author} {\bibinfo {author} {\bibfnamefont {J.}~\bibnamefont
  {Sanada}}, \bibinfo {author} {\bibfnamefont {M.}~\bibnamefont {Yamanouchi}},
  \bibinfo {author} {\bibfnamefont {N.}~\bibnamefont {Sakai}},\ and\ \bibinfo
  {author} {\bibfnamefont {S.}~\bibnamefont {Seki}},\ }\bibfield  {title}
  {\bibinfo {title} {Study of the excited states in {$^4$He} through the
  {$^4$He($\gamma, p$)$^3$H} reaction},\ }\href
  {https://doi.org/10.1143/JPSJ.28.537} {\bibfield  {journal} {\bibinfo
  {journal} {Journal of the Physical Society of Japan}\ }\textbf {\bibinfo
  {volume} {28}},\ \bibinfo {pages} {537} (\bibinfo {year} {1970})},\ \Eprint
  {https://arxiv.org/abs/https://doi.org/10.1143/JPSJ.28.537}
  {https://doi.org/10.1143/JPSJ.28.537} \BibitemShut {NoStop}%
\bibitem [{\citenamefont {Berman}\ \emph {et~al.}(1972)\citenamefont {Berman},
  \citenamefont {Firk},\ and\ \citenamefont {Wu}}]{Berman1972}%
  \BibitemOpen
  \bibfield  {author} {\bibinfo {author} {\bibfnamefont {B.}~\bibnamefont
  {Berman}}, \bibinfo {author} {\bibfnamefont {F.}~\bibnamefont {Firk}},\ and\
  \bibinfo {author} {\bibfnamefont {C.-P.}\ \bibnamefont {Wu}},\ }\bibfield
  {title} {\bibinfo {title} {The 90$\,^{\circ}$ differential cross section for
  the reaction {$^4$He ($\gamma, n_0$)$^3$He} and evidence for isospin mixing
  in the dipole states of {$^4$He}},\ }\href
  {https://doi.org/https://doi.org/10.1016/0375-9474(72)90620-3} {\bibfield
  {journal} {\bibinfo  {journal} {Nuclear Physics A}\ }\textbf {\bibinfo
  {volume} {179}},\ \bibinfo {pages} {791} (\bibinfo {year}
  {1972})}\BibitemShut {NoStop}%
\bibitem [{\citenamefont {Irish}\ \emph {et~al.}(1973)\citenamefont {Irish},
  \citenamefont {Johnson}, \citenamefont {Berman}, \citenamefont {Thomas},
  \citenamefont {McNeill},\ and\ \citenamefont {Jury}}]{Irish1973}%
  \BibitemOpen
  \bibfield  {author} {\bibinfo {author} {\bibfnamefont {J.~D.}\ \bibnamefont
  {Irish}}, \bibinfo {author} {\bibfnamefont {R.~G.}\ \bibnamefont {Johnson}},
  \bibinfo {author} {\bibfnamefont {B.~L.}\ \bibnamefont {Berman}}, \bibinfo
  {author} {\bibfnamefont {B.~J.}\ \bibnamefont {Thomas}}, \bibinfo {author}
  {\bibfnamefont {K.~G.}\ \bibnamefont {McNeill}},\ and\ \bibinfo {author}
  {\bibfnamefont {J.~W.}\ \bibnamefont {Jury}},\ }\bibfield  {title} {\bibinfo
  {title} {90\ifmmode^\circ\else\textdegree\fi{} differential cross section for
  the reaction $^{4}\mathrm{He}(\ensuremath{\gamma},n)^{3}\mathrm{He}$},\
  }\href {https://doi.org/10.1103/PhysRevC.8.1211} {\bibfield  {journal}
  {\bibinfo  {journal} {Phys. Rev. C}\ }\textbf {\bibinfo {volume} {8}},\
  \bibinfo {pages} {1211} (\bibinfo {year} {1973})}\BibitemShut {NoStop}%
\bibitem [{\citenamefont {Malcom}\ \emph {et~al.}(1973)\citenamefont {Malcom},
  \citenamefont {Webb}, \citenamefont {Shin},\ and\ \citenamefont
  {Skopik}}]{Malcom1973}%
  \BibitemOpen
  \bibfield  {author} {\bibinfo {author} {\bibfnamefont {C.}~\bibnamefont
  {Malcom}}, \bibinfo {author} {\bibfnamefont {D.}~\bibnamefont {Webb}},
  \bibinfo {author} {\bibfnamefont {Y.}~\bibnamefont {Shin}},\ and\ \bibinfo
  {author} {\bibfnamefont {D.}~\bibnamefont {Skopik}},\ }\bibfield  {title}
  {\bibinfo {title} {Evidence of a $2^+$ state from the
  {$^4$He($\gamma,n$)$^3$He} reaction},\ }\href
  {https://doi.org/https://doi.org/10.1016/0370-2693(73)90106-8} {\bibfield
  {journal} {\bibinfo  {journal} {Physics Letters B}\ }\textbf {\bibinfo
  {volume} {47}},\ \bibinfo {pages} {433} (\bibinfo {year} {1973})}\BibitemShut
  {NoStop}%
\bibitem [{\citenamefont {Arkatov}\ \emph {et~al.}(1974)\citenamefont
  {Arkatov}, \citenamefont {Vatset}, \citenamefont {Voloshchuk}, \citenamefont
  {Zolenko}, \citenamefont {Prokhorets},\ and\ \citenamefont
  {Chmil}}]{Arkatov1974}%
  \BibitemOpen
  \bibfield  {author} {\bibinfo {author} {\bibfnamefont {Y.~M.}\ \bibnamefont
  {Arkatov}}, \bibinfo {author} {\bibfnamefont {P.}~\bibnamefont {Vatset}},
  \bibinfo {author} {\bibfnamefont {V.}~\bibnamefont {Voloshchuk}}, \bibinfo
  {author} {\bibfnamefont {V.}~\bibnamefont {Zolenko}}, \bibinfo {author}
  {\bibfnamefont {I.}~\bibnamefont {Prokhorets}},\ and\ \bibinfo {author}
  {\bibfnamefont {V.}~\bibnamefont {Chmil}},\ }\bibfield  {title} {\bibinfo
  {title} {Energy moments of the total $\gamma$-quanta absorption cross section
  for {$^4$He} nucleus},\ }\href@noop {} {\bibfield  {journal} {\bibinfo
  {journal} {Yadernaya Fizika}\ }\textbf {\bibinfo {volume} {19}},\ \bibinfo
  {pages} {1172} (\bibinfo {year} {1974})},\ \bibinfo {note} {[Sov. J. nucl.
  Phys. \textbf{19}, 598 (1974)]}\BibitemShut {NoStop}%
\bibitem [{\citenamefont {Balestra}\ \emph {et~al.}(1977)\citenamefont
  {Balestra}, \citenamefont {Bollini}, \citenamefont {Busso}, \citenamefont
  {Garfagnini}, \citenamefont {Guaraldo}, \citenamefont {Piragino},
  \citenamefont {Scrimaglio},\ and\ \citenamefont {Zanini}}]{Balestra1977}%
  \BibitemOpen
  \bibfield  {author} {\bibinfo {author} {\bibfnamefont {F.}~\bibnamefont
  {Balestra}}, \bibinfo {author} {\bibfnamefont {E.}~\bibnamefont {Bollini}},
  \bibinfo {author} {\bibfnamefont {L.}~\bibnamefont {Busso}}, \bibinfo
  {author} {\bibfnamefont {R.}~\bibnamefont {Garfagnini}}, \bibinfo {author}
  {\bibfnamefont {C.}~\bibnamefont {Guaraldo}}, \bibinfo {author}
  {\bibfnamefont {G.}~\bibnamefont {Piragino}}, \bibinfo {author}
  {\bibfnamefont {R.}~\bibnamefont {Scrimaglio}},\ and\ \bibinfo {author}
  {\bibfnamefont {A.}~\bibnamefont {Zanini}},\ }\bibfield  {title} {\bibinfo
  {title} {Photodisintegration of {$^4$He} in the giant-resonance region},\
  }\href {https://doi.org/10.1007/BF02724538} {\bibfield  {journal} {\bibinfo
  {journal} {Il Nuovo Cimento A (1965-1970)}\ }\textbf {\bibinfo {volume}
  {38}},\ \bibinfo {pages} {145} (\bibinfo {year} {1977})}\BibitemShut
  {NoStop}%
\bibitem [{\citenamefont {Berman}\ \emph {et~al.}(1980)\citenamefont {Berman},
  \citenamefont {Faul}, \citenamefont {Meyer},\ and\ \citenamefont
  {Olson}}]{Berman1980}%
  \BibitemOpen
  \bibfield  {author} {\bibinfo {author} {\bibfnamefont {B.~L.}\ \bibnamefont
  {Berman}}, \bibinfo {author} {\bibfnamefont {D.~D.}\ \bibnamefont {Faul}},
  \bibinfo {author} {\bibfnamefont {P.}~\bibnamefont {Meyer}},\ and\ \bibinfo
  {author} {\bibfnamefont {D.~L.}\ \bibnamefont {Olson}},\ }\bibfield  {title}
  {\bibinfo {title} {Photoneutron cross section for $^{4}\mathrm{He}$},\ }\href
  {https://doi.org/10.1103/PhysRevC.22.2273} {\bibfield  {journal} {\bibinfo
  {journal} {Phys. Rev. C}\ }\textbf {\bibinfo {volume} {22}},\ \bibinfo
  {pages} {2273} (\bibinfo {year} {1980})}\BibitemShut {NoStop}%
\bibitem [{\citenamefont {Ward}\ \emph {et~al.}(1981)\citenamefont {Ward},
  \citenamefont {Tilley}, \citenamefont {Skopik}, \citenamefont {Roberson},\
  and\ \citenamefont {Weller}}]{Ward1981}%
  \BibitemOpen
  \bibfield  {author} {\bibinfo {author} {\bibfnamefont {L.}~\bibnamefont
  {Ward}}, \bibinfo {author} {\bibfnamefont {D.~R.}\ \bibnamefont {Tilley}},
  \bibinfo {author} {\bibfnamefont {D.~M.}\ \bibnamefont {Skopik}}, \bibinfo
  {author} {\bibfnamefont {N.~R.}\ \bibnamefont {Roberson}},\ and\ \bibinfo
  {author} {\bibfnamefont {H.~R.}\ \bibnamefont {Weller}},\ }\bibfield  {title}
  {\bibinfo {title} {Confirmation of the photoneutron cross section for
  $^{4}\mathrm{He}$ below 33 {MeV}},\ }\href
  {https://doi.org/10.1103/PhysRevC.24.317} {\bibfield  {journal} {\bibinfo
  {journal} {Phys. Rev. C}\ }\textbf {\bibinfo {volume} {24}},\ \bibinfo
  {pages} {317} (\bibinfo {year} {1981})}\BibitemShut {NoStop}%
\bibitem [{\citenamefont {McBroom}\ \emph {et~al.}(1982)\citenamefont
  {McBroom}, \citenamefont {Weller}, \citenamefont {Roberson},\ and\
  \citenamefont {Tilley}}]{McBroom1982}%
  \BibitemOpen
  \bibfield  {author} {\bibinfo {author} {\bibfnamefont {R.~C.}\ \bibnamefont
  {McBroom}}, \bibinfo {author} {\bibfnamefont {H.~R.}\ \bibnamefont {Weller}},
  \bibinfo {author} {\bibfnamefont {N.~R.}\ \bibnamefont {Roberson}},\ and\
  \bibinfo {author} {\bibfnamefont {D.~R.}\ \bibnamefont {Tilley}},\ }\bibfield
   {title} {\bibinfo {title}
  {$^{3}\mathrm{H}$($p$,$\ensuremath{\gamma}$)$^{4}\mathrm{He}$ reaction below
  ${E}_{p}=30$ {MeV}},\ }\href {https://doi.org/10.1103/PhysRevC.25.1644}
  {\bibfield  {journal} {\bibinfo  {journal} {Phys. Rev. C}\ }\textbf {\bibinfo
  {volume} {25}},\ \bibinfo {pages} {1644} (\bibinfo {year}
  {1982})}\BibitemShut {NoStop}%
\bibitem [{\citenamefont {Calarco}\ \emph {et~al.}(1983)\citenamefont
  {Calarco}, \citenamefont {Hanna}, \citenamefont {Chang}, \citenamefont
  {Diener}, \citenamefont {Kuhlmann},\ and\ \citenamefont
  {Fisher}}]{Calarco1983}%
  \BibitemOpen
  \bibfield  {author} {\bibinfo {author} {\bibfnamefont {J.~R.}\ \bibnamefont
  {Calarco}}, \bibinfo {author} {\bibfnamefont {S.~S.}\ \bibnamefont {Hanna}},
  \bibinfo {author} {\bibfnamefont {C.~C.}\ \bibnamefont {Chang}}, \bibinfo
  {author} {\bibfnamefont {E.~M.}\ \bibnamefont {Diener}}, \bibinfo {author}
  {\bibfnamefont {E.}~\bibnamefont {Kuhlmann}},\ and\ \bibinfo {author}
  {\bibfnamefont {G.~A.}\ \bibnamefont {Fisher}},\ }\bibfield  {title}
  {\bibinfo {title} {Absolute cross section for the reaction
  $^{3}\mathrm{H}(\mathrm{p}, {\ensuremath{\gamma}}_{0})^{4}\mathrm{He}$ and a
  review of $^{4}\mathrm{He}(\ensuremath{\gamma},
  {\mathrm{p}}_{0})^{3}\mathrm{H}$ measurements},\ }\href
  {https://doi.org/10.1103/PhysRevC.28.483} {\bibfield  {journal} {\bibinfo
  {journal} {Phys. Rev. C}\ }\textbf {\bibinfo {volume} {28}},\ \bibinfo
  {pages} {483} (\bibinfo {year} {1983})}\BibitemShut {NoStop}%
\bibitem [{\citenamefont {Bernabei}\ \emph {et~al.}(1988)\citenamefont
  {Bernabei}, \citenamefont {Chisholm}, \citenamefont {d'Angelo}, \citenamefont
  {De~Pascale}, \citenamefont {Picozza}, \citenamefont {Schaerf}, \citenamefont
  {Belli}, \citenamefont {Casano}, \citenamefont {Incicchitti}, \citenamefont
  {Prosperi},\ and\ \citenamefont {Girolami}}]{Bernabei1988}%
  \BibitemOpen
  \bibfield  {author} {\bibinfo {author} {\bibfnamefont {R.}~\bibnamefont
  {Bernabei}}, \bibinfo {author} {\bibfnamefont {A.}~\bibnamefont {Chisholm}},
  \bibinfo {author} {\bibfnamefont {S.}~\bibnamefont {d'Angelo}}, \bibinfo
  {author} {\bibfnamefont {M.~P.}\ \bibnamefont {De~Pascale}}, \bibinfo
  {author} {\bibfnamefont {P.}~\bibnamefont {Picozza}}, \bibinfo {author}
  {\bibfnamefont {C.}~\bibnamefont {Schaerf}}, \bibinfo {author} {\bibfnamefont
  {P.}~\bibnamefont {Belli}}, \bibinfo {author} {\bibfnamefont
  {L.}~\bibnamefont {Casano}}, \bibinfo {author} {\bibfnamefont
  {A.}~\bibnamefont {Incicchitti}}, \bibinfo {author} {\bibfnamefont
  {D.}~\bibnamefont {Prosperi}},\ and\ \bibinfo {author} {\bibfnamefont
  {B.}~\bibnamefont {Girolami}},\ }\bibfield  {title} {\bibinfo {title}
  {Measurement of the $^{4}\mathrm{He}$(\ensuremath{\gamma},$p$${)}^{3}${H}
  total cross section and charge symmetry},\ }\href
  {https://doi.org/10.1103/PhysRevC.38.1990} {\bibfield  {journal} {\bibinfo
  {journal} {Phys. Rev. C}\ }\textbf {\bibinfo {volume} {38}},\ \bibinfo
  {pages} {1990} (\bibinfo {year} {1988})}\BibitemShut {NoStop}%
\bibitem [{\citenamefont {Feldman}\ \emph {et~al.}(1990)\citenamefont
  {Feldman}, \citenamefont {Balbes}, \citenamefont {Kramer}, \citenamefont
  {Williams}, \citenamefont {Weller},\ and\ \citenamefont
  {Tilley}}]{Feldman1990}%
  \BibitemOpen
  \bibfield  {author} {\bibinfo {author} {\bibfnamefont {G.}~\bibnamefont
  {Feldman}}, \bibinfo {author} {\bibfnamefont {M.~J.}\ \bibnamefont {Balbes}},
  \bibinfo {author} {\bibfnamefont {L.~H.}\ \bibnamefont {Kramer}}, \bibinfo
  {author} {\bibfnamefont {J.~Z.}\ \bibnamefont {Williams}}, \bibinfo {author}
  {\bibfnamefont {H.~R.}\ \bibnamefont {Weller}},\ and\ \bibinfo {author}
  {\bibfnamefont {D.~R.}\ \bibnamefont {Tilley}},\ }\bibfield  {title}
  {\bibinfo {title} {$^{3}\mathrm{H}$($p$,\ensuremath{\gamma}${)}^{4}${He}
  reaction and the (\ensuremath{\gamma},$p$)/(\ensuremath{\gamma},$n$) ratio in
  $^{4}\mathrm{He}$},\ }\href {https://doi.org/10.1103/PhysRevC.42.R1167}
  {\bibfield  {journal} {\bibinfo  {journal} {Phys. Rev. C}\ }\textbf {\bibinfo
  {volume} {42}},\ \bibinfo {pages} {R1167} (\bibinfo {year}
  {1990})}\BibitemShut {NoStop}%
\bibitem [{\citenamefont {Komar}\ \emph {et~al.}(1993)\citenamefont {Komar},
  \citenamefont {Mak}, \citenamefont {Leslie}, \citenamefont {Evans},
  \citenamefont {Bonvin}, \citenamefont {Earle},\ and\ \citenamefont
  {Alexander}}]{Komer1993}%
  \BibitemOpen
  \bibfield  {author} {\bibinfo {author} {\bibfnamefont {R.~J.}\ \bibnamefont
  {Komar}}, \bibinfo {author} {\bibfnamefont {H.-B.}\ \bibnamefont {Mak}},
  \bibinfo {author} {\bibfnamefont {J.~R.}\ \bibnamefont {Leslie}}, \bibinfo
  {author} {\bibfnamefont {H.~C.}\ \bibnamefont {Evans}}, \bibinfo {author}
  {\bibfnamefont {E.}~\bibnamefont {Bonvin}}, \bibinfo {author} {\bibfnamefont
  {E.~D.}\ \bibnamefont {Earle}},\ and\ \bibinfo {author} {\bibfnamefont
  {T.~K.}\ \bibnamefont {Alexander}},\ }\bibfield  {title} {\bibinfo {title}
  {$^{3}\mathrm{He}$(n,\ensuremath{\gamma}${)}^{4}${He} cross section and the
  photodisintegration of $^{4}\mathrm{He}$},\ }\href
  {https://doi.org/10.1103/PhysRevC.48.2375} {\bibfield  {journal} {\bibinfo
  {journal} {Phys. Rev. C}\ }\textbf {\bibinfo {volume} {48}},\ \bibinfo
  {pages} {2375} (\bibinfo {year} {1993})}\BibitemShut {NoStop}%
\bibitem [{\citenamefont {Van~Hoorebeke}\ \emph {et~al.}(1993)\citenamefont
  {Van~Hoorebeke}, \citenamefont {Van~de Vyver}, \citenamefont {Fiermans},
  \citenamefont {Ryckbosch}, \citenamefont {Van~den Abeele},\ and\
  \citenamefont {Dias}}]{Hoorebeke1993}%
  \BibitemOpen
  \bibfield  {author} {\bibinfo {author} {\bibfnamefont {L.}~\bibnamefont
  {Van~Hoorebeke}}, \bibinfo {author} {\bibfnamefont {R.}~\bibnamefont {Van~de
  Vyver}}, \bibinfo {author} {\bibfnamefont {V.}~\bibnamefont {Fiermans}},
  \bibinfo {author} {\bibfnamefont {D.}~\bibnamefont {Ryckbosch}}, \bibinfo
  {author} {\bibfnamefont {C.}~\bibnamefont {Van~den Abeele}},\ and\ \bibinfo
  {author} {\bibfnamefont {J.}~\bibnamefont {Dias}},\ }\bibfield  {title}
  {\bibinfo {title} {Direct measurement of the
  $^{4}\mathrm{He}$(\ensuremath{\gamma},${\mathit{p}}_{0}$) absolute cross
  section},\ }\href {https://doi.org/10.1103/PhysRevC.48.2510} {\bibfield
  {journal} {\bibinfo  {journal} {Phys. Rev. C}\ }\textbf {\bibinfo {volume}
  {48}},\ \bibinfo {pages} {2510} (\bibinfo {year} {1993})}\BibitemShut
  {NoStop}%
\bibitem [{\citenamefont {Hahn}\ \emph {et~al.}(1995)\citenamefont {Hahn},
  \citenamefont {Brune},\ and\ \citenamefont {Kavanagh}}]{Hahn1995}%
  \BibitemOpen
  \bibfield  {author} {\bibinfo {author} {\bibfnamefont {K.~I.}\ \bibnamefont
  {Hahn}}, \bibinfo {author} {\bibfnamefont {C.~R.}\ \bibnamefont {Brune}},\
  and\ \bibinfo {author} {\bibfnamefont {R.~W.}\ \bibnamefont {Kavanagh}},\
  }\bibfield  {title} {\bibinfo {title}
  {$^{3}\mathrm{H}$($p$,\ensuremath{\gamma}${)}^{4}${He} cross section},\
  }\href {https://doi.org/10.1103/PhysRevC.51.1624} {\bibfield  {journal}
  {\bibinfo  {journal} {Phys. Rev. C}\ }\textbf {\bibinfo {volume} {51}},\
  \bibinfo {pages} {1624} (\bibinfo {year} {1995})}\BibitemShut {NoStop}%
\bibitem [{\citenamefont {Shima}\ \emph {et~al.}(2005)\citenamefont {Shima},
  \citenamefont {Naito}, \citenamefont {Nagai}, \citenamefont {Baba},
  \citenamefont {Tamura}, \citenamefont {Takahashi}, \citenamefont {Kii},
  \citenamefont {Ohgaki},\ and\ \citenamefont {Toyokawa}}]{Shima2005}%
  \BibitemOpen
  \bibfield  {author} {\bibinfo {author} {\bibfnamefont {T.}~\bibnamefont
  {Shima}}, \bibinfo {author} {\bibfnamefont {S.}~\bibnamefont {Naito}},
  \bibinfo {author} {\bibfnamefont {Y.}~\bibnamefont {Nagai}}, \bibinfo
  {author} {\bibfnamefont {T.}~\bibnamefont {Baba}}, \bibinfo {author}
  {\bibfnamefont {K.}~\bibnamefont {Tamura}}, \bibinfo {author} {\bibfnamefont
  {T.}~\bibnamefont {Takahashi}}, \bibinfo {author} {\bibfnamefont
  {T.}~\bibnamefont {Kii}}, \bibinfo {author} {\bibfnamefont {H.}~\bibnamefont
  {Ohgaki}},\ and\ \bibinfo {author} {\bibfnamefont {H.}~\bibnamefont
  {Toyokawa}},\ }\bibfield  {title} {\bibinfo {title} {Simultaneous measurement
  of the photodisintegration of $^{4}\mathrm{He}$ in the giant dipole resonance
  region},\ }\href {https://doi.org/10.1103/PhysRevC.72.044004} {\bibfield
  {journal} {\bibinfo  {journal} {Phys. Rev. C}\ }\textbf {\bibinfo {volume}
  {72}},\ \bibinfo {pages} {044004} (\bibinfo {year} {2005})}\BibitemShut
  {NoStop}%
\bibitem [{\citenamefont {Nilsson}\ \emph {et~al.}(2007)\citenamefont
  {Nilsson}, \citenamefont {Adler}, \citenamefont {Andersson}, \citenamefont
  {Annand}, \citenamefont {Akkurt}, \citenamefont {Boland}, \citenamefont
  {Crawford}, \citenamefont {Fissum}, \citenamefont {Hansen}, \citenamefont
  {Harty}, \citenamefont {Ireland}, \citenamefont {Isaksson}, \citenamefont
  {Karlsson}, \citenamefont {Lundin}, \citenamefont {McGeorge}, \citenamefont
  {Miller}, \citenamefont {Ruijter}, \citenamefont {Sandell}, \citenamefont
  {Schr\"oder}, \citenamefont {Sims},\ and\ \citenamefont
  {Watts}}]{Nilsson2007}%
  \BibitemOpen
  \bibfield  {author} {\bibinfo {author} {\bibfnamefont {B.}~\bibnamefont
  {Nilsson}}, \bibinfo {author} {\bibfnamefont {J.-O.}\ \bibnamefont {Adler}},
  \bibinfo {author} {\bibfnamefont {B.-E.}\ \bibnamefont {Andersson}}, \bibinfo
  {author} {\bibfnamefont {J.~R.~M.}\ \bibnamefont {Annand}}, \bibinfo {author}
  {\bibfnamefont {I.}~\bibnamefont {Akkurt}}, \bibinfo {author} {\bibfnamefont
  {M.~J.}\ \bibnamefont {Boland}}, \bibinfo {author} {\bibfnamefont {G.~I.}\
  \bibnamefont {Crawford}}, \bibinfo {author} {\bibfnamefont {K.~G.}\
  \bibnamefont {Fissum}}, \bibinfo {author} {\bibfnamefont {K.}~\bibnamefont
  {Hansen}}, \bibinfo {author} {\bibfnamefont {P.~D.}\ \bibnamefont {Harty}},
  \bibinfo {author} {\bibfnamefont {D.~G.}\ \bibnamefont {Ireland}}, \bibinfo
  {author} {\bibfnamefont {L.}~\bibnamefont {Isaksson}}, \bibinfo {author}
  {\bibfnamefont {M.}~\bibnamefont {Karlsson}}, \bibinfo {author}
  {\bibfnamefont {M.}~\bibnamefont {Lundin}}, \bibinfo {author} {\bibfnamefont
  {J.~C.}\ \bibnamefont {McGeorge}}, \bibinfo {author} {\bibfnamefont {G.~J.}\
  \bibnamefont {Miller}}, \bibinfo {author} {\bibfnamefont {H.}~\bibnamefont
  {Ruijter}}, \bibinfo {author} {\bibfnamefont {A.}~\bibnamefont {Sandell}},
  \bibinfo {author} {\bibfnamefont {B.}~\bibnamefont {Schr\"oder}}, \bibinfo
  {author} {\bibfnamefont {D.~A.}\ \bibnamefont {Sims}},\ and\ \bibinfo
  {author} {\bibfnamefont {D.}~\bibnamefont {Watts}} (\bibinfo {collaboration}
  {The MAX-lab Nuclear Physics Working Group}),\ }\bibfield  {title} {\bibinfo
  {title} {Measurement of the $^{4}\mathrm{He}$$(\ensuremath{\gamma},n)$
  reaction from $23<{E}_{\ensuremath{\gamma}}<70$ {MeV}},\ }\href
  {https://doi.org/10.1103/PhysRevC.75.014007} {\bibfield  {journal} {\bibinfo
  {journal} {Phys. Rev. C}\ }\textbf {\bibinfo {volume} {75}},\ \bibinfo
  {pages} {014007} (\bibinfo {year} {2007})}\BibitemShut {NoStop}%
\bibitem [{\citenamefont {Nakayama}\ \emph {et~al.}(2007)\citenamefont
  {Nakayama}, \citenamefont {Matsumoto}, \citenamefont {Hayami}, \citenamefont
  {Fushimi}, \citenamefont {Kawasuso}, \citenamefont {Yasuda}, \citenamefont
  {Yamagata}, \citenamefont {Akimune}, \citenamefont {Ikemizu}, \citenamefont
  {Fujiwara}, \citenamefont {Yosoi}, \citenamefont {Nakanishi}, \citenamefont
  {Kawase}, \citenamefont {Hashimoto}, \citenamefont {Oota}, \citenamefont
  {Sagara}, \citenamefont {Kudoh}, \citenamefont {Asaji}, \citenamefont
  {Ishida}, \citenamefont {Tanaka},\ and\ \citenamefont
  {Greenfield}}]{Nakayama2007}%
  \BibitemOpen
  \bibfield  {author} {\bibinfo {author} {\bibfnamefont {S.}~\bibnamefont
  {Nakayama}}, \bibinfo {author} {\bibfnamefont {E.}~\bibnamefont {Matsumoto}},
  \bibinfo {author} {\bibfnamefont {R.}~\bibnamefont {Hayami}}, \bibinfo
  {author} {\bibfnamefont {K.}~\bibnamefont {Fushimi}}, \bibinfo {author}
  {\bibfnamefont {H.}~\bibnamefont {Kawasuso}}, \bibinfo {author}
  {\bibfnamefont {K.}~\bibnamefont {Yasuda}}, \bibinfo {author} {\bibfnamefont
  {T.}~\bibnamefont {Yamagata}}, \bibinfo {author} {\bibfnamefont
  {H.}~\bibnamefont {Akimune}}, \bibinfo {author} {\bibfnamefont
  {H.}~\bibnamefont {Ikemizu}}, \bibinfo {author} {\bibfnamefont
  {M.}~\bibnamefont {Fujiwara}}, \bibinfo {author} {\bibfnamefont
  {M.}~\bibnamefont {Yosoi}}, \bibinfo {author} {\bibfnamefont
  {K.}~\bibnamefont {Nakanishi}}, \bibinfo {author} {\bibfnamefont
  {K.}~\bibnamefont {Kawase}}, \bibinfo {author} {\bibfnamefont
  {H.}~\bibnamefont {Hashimoto}}, \bibinfo {author} {\bibfnamefont
  {T.}~\bibnamefont {Oota}}, \bibinfo {author} {\bibfnamefont {K.}~\bibnamefont
  {Sagara}}, \bibinfo {author} {\bibfnamefont {T.}~\bibnamefont {Kudoh}},
  \bibinfo {author} {\bibfnamefont {S.}~\bibnamefont {Asaji}}, \bibinfo
  {author} {\bibfnamefont {T.}~\bibnamefont {Ishida}}, \bibinfo {author}
  {\bibfnamefont {M.}~\bibnamefont {Tanaka}},\ and\ \bibinfo {author}
  {\bibfnamefont {M.~B.}\ \bibnamefont {Greenfield}},\ }\bibfield  {title}
  {\bibinfo {title} {Analog of the giant dipole resonance in
  $^{4}\mathrm{He}$},\ }\href {https://doi.org/10.1103/PhysRevC.76.021305}
  {\bibfield  {journal} {\bibinfo  {journal} {Phys. Rev. C}\ }\textbf {\bibinfo
  {volume} {76}},\ \bibinfo {pages} {021305(R)} (\bibinfo {year}
  {2007})}\BibitemShut {NoStop}%
\bibitem [{\citenamefont {Shima}\ \emph {et~al.}(2010)\citenamefont {Shima},
  \citenamefont {Nagai}, \citenamefont {Miyamoto}, \citenamefont {Amano},
  \citenamefont {Horikawa},\ and\ \citenamefont {Mochizuki}}]{Shima2010}%
  \BibitemOpen
  \bibfield  {author} {\bibinfo {author} {\bibfnamefont {T.}~\bibnamefont
  {Shima}}, \bibinfo {author} {\bibfnamefont {Y.}~\bibnamefont {Nagai}},
  \bibinfo {author} {\bibfnamefont {S.}~\bibnamefont {Miyamoto}}, \bibinfo
  {author} {\bibfnamefont {S.}~\bibnamefont {Amano}}, \bibinfo {author}
  {\bibfnamefont {K.}~\bibnamefont {Horikawa}},\ and\ \bibinfo {author}
  {\bibfnamefont {T.}~\bibnamefont {Mochizuki}},\ }\bibfield  {title} {\bibinfo
  {title} {{New results on photodisintegration of $^4$He}},\ }\href
  {https://doi.org/10.22323/1.086.0107} {\bibfield  {journal} {\bibinfo
  {journal} {PoS}\ }\textbf {\bibinfo {volume} {CD09}},\ \bibinfo {pages} {107}
  (\bibinfo {year} {2010})}\BibitemShut {NoStop}%
\bibitem [{\citenamefont {Tornow}\ \emph {et~al.}(2012)\citenamefont {Tornow},
  \citenamefont {Kelley}, \citenamefont {Raut}, \citenamefont {Rusev},
  \citenamefont {Tonchev}, \citenamefont {Ahmed}, \citenamefont {Crowell},\
  and\ \citenamefont {Stave}}]{Tornow2012}%
  \BibitemOpen
  \bibfield  {author} {\bibinfo {author} {\bibfnamefont {W.}~\bibnamefont
  {Tornow}}, \bibinfo {author} {\bibfnamefont {J.~H.}\ \bibnamefont {Kelley}},
  \bibinfo {author} {\bibfnamefont {R.}~\bibnamefont {Raut}}, \bibinfo {author}
  {\bibfnamefont {G.}~\bibnamefont {Rusev}}, \bibinfo {author} {\bibfnamefont
  {A.~P.}\ \bibnamefont {Tonchev}}, \bibinfo {author} {\bibfnamefont {M.~W.}\
  \bibnamefont {Ahmed}}, \bibinfo {author} {\bibfnamefont {A.~S.}\ \bibnamefont
  {Crowell}},\ and\ \bibinfo {author} {\bibfnamefont {S.~C.}\ \bibnamefont
  {Stave}},\ }\bibfield  {title} {\bibinfo {title} {Photodisintegration cross
  section of the reaction ${}^{4}${He}($\ensuremath{\gamma}$,$n$)${}^{3}${He}
  at the giant dipole resonance peak},\ }\href
  {https://doi.org/10.1103/PhysRevC.85.061001} {\bibfield  {journal} {\bibinfo
  {journal} {Phys. Rev. C}\ }\textbf {\bibinfo {volume} {85}},\ \bibinfo
  {pages} {061001(R)} (\bibinfo {year} {2012})}\BibitemShut {NoStop}%
\bibitem [{\citenamefont {Raut}\ \emph {et~al.}(2012)\citenamefont {Raut},
  \citenamefont {Tornow}, \citenamefont {Ahmed}, \citenamefont {Crowell},
  \citenamefont {Kelley}, \citenamefont {Rusev}, \citenamefont {Stave},\ and\
  \citenamefont {Tonchev}}]{Raut2012}%
  \BibitemOpen
  \bibfield  {author} {\bibinfo {author} {\bibfnamefont {R.}~\bibnamefont
  {Raut}}, \bibinfo {author} {\bibfnamefont {W.}~\bibnamefont {Tornow}},
  \bibinfo {author} {\bibfnamefont {M.~W.}\ \bibnamefont {Ahmed}}, \bibinfo
  {author} {\bibfnamefont {A.~S.}\ \bibnamefont {Crowell}}, \bibinfo {author}
  {\bibfnamefont {J.~H.}\ \bibnamefont {Kelley}}, \bibinfo {author}
  {\bibfnamefont {G.}~\bibnamefont {Rusev}}, \bibinfo {author} {\bibfnamefont
  {S.~C.}\ \bibnamefont {Stave}},\ and\ \bibinfo {author} {\bibfnamefont
  {A.~P.}\ \bibnamefont {Tonchev}},\ }\bibfield  {title} {\bibinfo {title}
  {Photodisintegration cross section of the reaction
  $^{4}\mathrm{He}(\ensuremath{\gamma},$p$)^{3}\mathrm{H}$ between 22 and 30
  {MeV}},\ }\href {https://doi.org/10.1103/PhysRevLett.108.042502} {\bibfield
  {journal} {\bibinfo  {journal} {Phys. Rev. Lett.}\ }\textbf {\bibinfo
  {volume} {108}},\ \bibinfo {pages} {042502} (\bibinfo {year}
  {2012})}\BibitemShut {NoStop}%
\bibitem [{\citenamefont {Quaglioni}\ \emph {et~al.}(2004)\citenamefont
  {Quaglioni}, \citenamefont {Leidemann}, \citenamefont {Orlandini},
  \citenamefont {Barnea},\ and\ \citenamefont {Efros}}]{Quaglioni2004}%
  \BibitemOpen
  \bibfield  {author} {\bibinfo {author} {\bibfnamefont {S.}~\bibnamefont
  {Quaglioni}}, \bibinfo {author} {\bibfnamefont {W.}~\bibnamefont
  {Leidemann}}, \bibinfo {author} {\bibfnamefont {G.}~\bibnamefont
  {Orlandini}}, \bibinfo {author} {\bibfnamefont {N.}~\bibnamefont {Barnea}},\
  and\ \bibinfo {author} {\bibfnamefont {V.~D.}\ \bibnamefont {Efros}},\
  }\bibfield  {title} {\bibinfo {title} {Two-body photodisintegration of
  $^{4}\mathrm{He}$ with full final state interaction},\ }\href
  {https://doi.org/10.1103/PhysRevC.69.044002} {\bibfield  {journal} {\bibinfo
  {journal} {Phys. Rev. C}\ }\textbf {\bibinfo {volume} {69}},\ \bibinfo
  {pages} {044002} (\bibinfo {year} {2004})}\BibitemShut {NoStop}%
\bibitem [{\citenamefont {Horiuchi}\ \emph {et~al.}(2012)\citenamefont
  {Horiuchi}, \citenamefont {Suzuki},\ and\ \citenamefont
  {Arai}}]{Horiuchi2012}%
  \BibitemOpen
  \bibfield  {author} {\bibinfo {author} {\bibfnamefont {W.}~\bibnamefont
  {Horiuchi}}, \bibinfo {author} {\bibfnamefont {Y.}~\bibnamefont {Suzuki}},\
  and\ \bibinfo {author} {\bibfnamefont {K.}~\bibnamefont {Arai}},\ }\bibfield
  {title} {\bibinfo {title} {Ab initio study of the photoabsorption of
  ${}^{4}${He}},\ }\href {https://doi.org/10.1103/PhysRevC.85.054002}
  {\bibfield  {journal} {\bibinfo  {journal} {Phys. Rev. C}\ }\textbf {\bibinfo
  {volume} {85}},\ \bibinfo {pages} {054002} (\bibinfo {year}
  {2012})}\BibitemShut {NoStop}%
\bibitem [{\citenamefont {Miyamoto}\ \emph {et~al.}(2006)\citenamefont
  {Miyamoto}, \citenamefont {Asano}, \citenamefont {Amano}, \citenamefont {Li},
  \citenamefont {Imasaki}, \citenamefont {Kinugasa}, \citenamefont {Shoji},
  \citenamefont {Takagi},\ and\ \citenamefont {Mochizuki}}]{Miyamoto2006}%
  \BibitemOpen
  \bibfield  {author} {\bibinfo {author} {\bibfnamefont {S.}~\bibnamefont
  {Miyamoto}}, \bibinfo {author} {\bibfnamefont {Y.}~\bibnamefont {Asano}},
  \bibinfo {author} {\bibfnamefont {S.}~\bibnamefont {Amano}}, \bibinfo
  {author} {\bibfnamefont {D.}~\bibnamefont {Li}}, \bibinfo {author}
  {\bibfnamefont {K.}~\bibnamefont {Imasaki}}, \bibinfo {author} {\bibfnamefont
  {H.}~\bibnamefont {Kinugasa}}, \bibinfo {author} {\bibfnamefont
  {Y.}~\bibnamefont {Shoji}}, \bibinfo {author} {\bibfnamefont
  {T.}~\bibnamefont {Takagi}},\ and\ \bibinfo {author} {\bibfnamefont
  {T.}~\bibnamefont {Mochizuki}},\ }\bibfield  {title} {\bibinfo {title} {Laser
  compton back-scattering gamma-ray beamline on newsubaru},\ }\href
  {https://doi.org/https://doi.org/10.1016/j.radmeas.2007.01.013} {\bibfield
  {journal} {\bibinfo  {journal} {Radiation Measurements}\ }\textbf {\bibinfo
  {volume} {41}},\ \bibinfo {pages} {S179} (\bibinfo {year} {2006})},\ \bibinfo
  {note} {the 3rd International Workshop on Radiation Safety at Synchrotron
  Radiation Sources}\BibitemShut {NoStop}%
\bibitem [{\citenamefont {Utsunomiya}\ \emph {et~al.}(2014)\citenamefont
  {Utsunomiya}, \citenamefont {Shima}, \citenamefont {Takahisa}, \citenamefont
  {Filipescu}, \citenamefont {Tesileanu}, \citenamefont {Gheorghe},
  \citenamefont {Nyhus}, \citenamefont {Renstr^^c3^^b8m}, \citenamefont {Lui},
  \citenamefont {Kitagawa}, \citenamefont {Amano},\ and\ \citenamefont
  {Miyamoto}}]{Utsunomiya2014}%
  \BibitemOpen
  \bibfield  {author} {\bibinfo {author} {\bibfnamefont {H.}~\bibnamefont
  {Utsunomiya}}, \bibinfo {author} {\bibfnamefont {T.}~\bibnamefont {Shima}},
  \bibinfo {author} {\bibfnamefont {K.}~\bibnamefont {Takahisa}}, \bibinfo
  {author} {\bibfnamefont {D.~M.}\ \bibnamefont {Filipescu}}, \bibinfo {author}
  {\bibfnamefont {O.}~\bibnamefont {Tesileanu}}, \bibinfo {author}
  {\bibfnamefont {I.}~\bibnamefont {Gheorghe}}, \bibinfo {author}
  {\bibfnamefont {H.-T.}\ \bibnamefont {Nyhus}}, \bibinfo {author}
  {\bibfnamefont {T.}~\bibnamefont {Renstr^^c3^^b8m}}, \bibinfo {author}
  {\bibfnamefont {Y.-W.}\ \bibnamefont {Lui}}, \bibinfo {author} {\bibfnamefont
  {Y.}~\bibnamefont {Kitagawa}}, \bibinfo {author} {\bibfnamefont
  {S.}~\bibnamefont {Amano}},\ and\ \bibinfo {author} {\bibfnamefont
  {S.}~\bibnamefont {Miyamoto}},\ }\bibfield  {title} {\bibinfo {title} {Energy
  calibration of the newsubaru storage ring for laser compton-scattering gamma
  rays and applications},\ }\href {https://doi.org/10.1109/TNS.2014.2312323}
  {\bibfield  {journal} {\bibinfo  {journal} {IEEE Transactions on Nuclear
  Science}\ }\textbf {\bibinfo {volume} {61}},\ \bibinfo {pages} {1252}
  (\bibinfo {year} {2014})}\BibitemShut {NoStop}%
\bibitem [{\citenamefont {D'angelo}\ \emph {et~al.}(2000)\citenamefont
  {D'angelo}, \citenamefont {Bartalini}, \citenamefont {Bellini}, \citenamefont
  {Sandri}, \citenamefont {Moricciani}, \citenamefont {Nicoletti},\ and\
  \citenamefont {Zucchiatti}}]{D'angelo2000}%
  \BibitemOpen
  \bibfield  {author} {\bibinfo {author} {\bibfnamefont {A.}~\bibnamefont
  {D'angelo}}, \bibinfo {author} {\bibfnamefont {O.}~\bibnamefont {Bartalini}},
  \bibinfo {author} {\bibfnamefont {V.}~\bibnamefont {Bellini}}, \bibinfo
  {author} {\bibfnamefont {P.~L.}\ \bibnamefont {Sandri}}, \bibinfo {author}
  {\bibfnamefont {D.}~\bibnamefont {Moricciani}}, \bibinfo {author}
  {\bibfnamefont {L.}~\bibnamefont {Nicoletti}},\ and\ \bibinfo {author}
  {\bibfnamefont {A.}~\bibnamefont {Zucchiatti}},\ }\bibfield  {title}
  {\bibinfo {title} {Generation of compton backscattering-ray beams},\
  }\href@noop {} {\bibfield  {journal} {\bibinfo  {journal} {Nuclear
  Instruments and Methods in Physics Research A}\ }\textbf {\bibinfo {volume}
  {455}},\ \bibinfo {pages} {1} (\bibinfo {year} {2000})}\BibitemShut {NoStop}%
\bibitem [{\citenamefont {Furuno}\ \emph {et~al.}(2018)\citenamefont {Furuno},
  \citenamefont {Kawabata}, \citenamefont {Ong}, \citenamefont {Adachi},
  \citenamefont {Ayyad}, \citenamefont {Baba}, \citenamefont {Fujikawa},
  \citenamefont {Hashimoto}, \citenamefont {Inaba}, \citenamefont {Ishii},
  \citenamefont {Kabuki}, \citenamefont {Kubo}, \citenamefont {Matsuda},
  \citenamefont {Matsuoka}, \citenamefont {Mizumoto}, \citenamefont {Morimoto},
  \citenamefont {Murata}, \citenamefont {Sawano}, \citenamefont {Suzuki},
  \citenamefont {Takada}, \citenamefont {Tanaka}, \citenamefont {Tanihata},
  \citenamefont {Tanimori}, \citenamefont {Tran}, \citenamefont {Tsumura},\
  and\ \citenamefont {Watanabe}}]{Furuno2018}%
  \BibitemOpen
  \bibfield  {author} {\bibinfo {author} {\bibfnamefont {T.}~\bibnamefont
  {Furuno}}, \bibinfo {author} {\bibfnamefont {T.}~\bibnamefont {Kawabata}},
  \bibinfo {author} {\bibfnamefont {H.}~\bibnamefont {Ong}}, \bibinfo {author}
  {\bibfnamefont {S.}~\bibnamefont {Adachi}}, \bibinfo {author} {\bibfnamefont
  {Y.}~\bibnamefont {Ayyad}}, \bibinfo {author} {\bibfnamefont
  {T.}~\bibnamefont {Baba}}, \bibinfo {author} {\bibfnamefont {Y.}~\bibnamefont
  {Fujikawa}}, \bibinfo {author} {\bibfnamefont {T.}~\bibnamefont {Hashimoto}},
  \bibinfo {author} {\bibfnamefont {K.}~\bibnamefont {Inaba}}, \bibinfo
  {author} {\bibfnamefont {Y.}~\bibnamefont {Ishii}}, \bibinfo {author}
  {\bibfnamefont {S.}~\bibnamefont {Kabuki}}, \bibinfo {author} {\bibfnamefont
  {H.}~\bibnamefont {Kubo}}, \bibinfo {author} {\bibfnamefont {Y.}~\bibnamefont
  {Matsuda}}, \bibinfo {author} {\bibfnamefont {Y.}~\bibnamefont {Matsuoka}},
  \bibinfo {author} {\bibfnamefont {T.}~\bibnamefont {Mizumoto}}, \bibinfo
  {author} {\bibfnamefont {T.}~\bibnamefont {Morimoto}}, \bibinfo {author}
  {\bibfnamefont {M.}~\bibnamefont {Murata}}, \bibinfo {author} {\bibfnamefont
  {T.}~\bibnamefont {Sawano}}, \bibinfo {author} {\bibfnamefont
  {T.}~\bibnamefont {Suzuki}}, \bibinfo {author} {\bibfnamefont
  {A.}~\bibnamefont {Takada}}, \bibinfo {author} {\bibfnamefont
  {J.}~\bibnamefont {Tanaka}}, \bibinfo {author} {\bibfnamefont
  {I.}~\bibnamefont {Tanihata}}, \bibinfo {author} {\bibfnamefont
  {T.}~\bibnamefont {Tanimori}}, \bibinfo {author} {\bibfnamefont
  {D.}~\bibnamefont {Tran}}, \bibinfo {author} {\bibfnamefont {M.}~\bibnamefont
  {Tsumura}},\ and\ \bibinfo {author} {\bibfnamefont {H.}~\bibnamefont
  {Watanabe}},\ }\bibfield  {title} {\bibinfo {title} {Performance test of the
  {MAIKo} active target},\ }\href
  {https://doi.org/https://doi.org/10.1016/j.nima.2018.08.042} {\bibfield
  {journal} {\bibinfo  {journal} {Nuclear Instruments and Methods in Physics
  Research Section A: Accelerators, Spectrometers, Detectors and Associated
  Equipment}\ }\textbf {\bibinfo {volume} {908}},\ \bibinfo {pages} {215}
  (\bibinfo {year} {2018})}\BibitemShut {NoStop}%
\bibitem [{\citenamefont {Ochi}\ \emph {et~al.}(2002)\citenamefont {Ochi},
  \citenamefont {Nagayoshi}, \citenamefont {Koishi}, \citenamefont {Tanimori},
  \citenamefont {Nagae},\ and\ \citenamefont {Nakamura}}]{Ochi2002}%
  \BibitemOpen
  \bibfield  {author} {\bibinfo {author} {\bibfnamefont {A.}~\bibnamefont
  {Ochi}}, \bibinfo {author} {\bibfnamefont {T.}~\bibnamefont {Nagayoshi}},
  \bibinfo {author} {\bibfnamefont {S.}~\bibnamefont {Koishi}}, \bibinfo
  {author} {\bibfnamefont {T.}~\bibnamefont {Tanimori}}, \bibinfo {author}
  {\bibfnamefont {T.}~\bibnamefont {Nagae}},\ and\ \bibinfo {author}
  {\bibfnamefont {M.}~\bibnamefont {Nakamura}},\ }\bibfield  {title} {\bibinfo
  {title} {Development of micro pixel chamber},\ }\href
  {https://doi.org/https://doi.org/10.1016/S0168-9002(01)01756-9} {\bibfield
  {journal} {\bibinfo  {journal} {Nuclear Instruments and Methods in Physics
  Research Section A: Accelerators, Spectrometers, Detectors and Associated
  Equipment}\ }\textbf {\bibinfo {volume} {478}},\ \bibinfo {pages} {196}
  (\bibinfo {year} {2002})},\ \bibinfo {note} {proceedings of the ninth
  Int.Conf. on Instrumentation}\BibitemShut {NoStop}%
\bibitem [{\citenamefont {Mizumoto}\ \emph {et~al.}(2015)\citenamefont
  {Mizumoto}, \citenamefont {Matsuoka}, \citenamefont {Mizumura}, \citenamefont
  {Tanimori}, \citenamefont {Kubo}, \citenamefont {Takada}, \citenamefont
  {Iwaki}, \citenamefont {Sawano}, \citenamefont {Nakamura}, \citenamefont
  {Komura}, \citenamefont {Nakamura}, \citenamefont {Kishimoto}, \citenamefont
  {Oda}, \citenamefont {Miyamoto}, \citenamefont {Takemura}, \citenamefont
  {Parker}, \citenamefont {Tomono}, \citenamefont {Sonoda}, \citenamefont
  {Miuchi},\ and\ \citenamefont {Kurosawa}}]{Mizumoto2015}%
  \BibitemOpen
  \bibfield  {author} {\bibinfo {author} {\bibfnamefont {T.}~\bibnamefont
  {Mizumoto}}, \bibinfo {author} {\bibfnamefont {Y.}~\bibnamefont {Matsuoka}},
  \bibinfo {author} {\bibfnamefont {Y.}~\bibnamefont {Mizumura}}, \bibinfo
  {author} {\bibfnamefont {T.}~\bibnamefont {Tanimori}}, \bibinfo {author}
  {\bibfnamefont {H.}~\bibnamefont {Kubo}}, \bibinfo {author} {\bibfnamefont
  {A.}~\bibnamefont {Takada}}, \bibinfo {author} {\bibfnamefont
  {S.}~\bibnamefont {Iwaki}}, \bibinfo {author} {\bibfnamefont
  {T.}~\bibnamefont {Sawano}}, \bibinfo {author} {\bibfnamefont
  {K.}~\bibnamefont {Nakamura}}, \bibinfo {author} {\bibfnamefont
  {S.}~\bibnamefont {Komura}}, \bibinfo {author} {\bibfnamefont
  {S.}~\bibnamefont {Nakamura}}, \bibinfo {author} {\bibfnamefont
  {T.}~\bibnamefont {Kishimoto}}, \bibinfo {author} {\bibfnamefont
  {M.}~\bibnamefont {Oda}}, \bibinfo {author} {\bibfnamefont {S.}~\bibnamefont
  {Miyamoto}}, \bibinfo {author} {\bibfnamefont {T.}~\bibnamefont {Takemura}},
  \bibinfo {author} {\bibfnamefont {J.~D.}\ \bibnamefont {Parker}}, \bibinfo
  {author} {\bibfnamefont {D.}~\bibnamefont {Tomono}}, \bibinfo {author}
  {\bibfnamefont {S.}~\bibnamefont {Sonoda}}, \bibinfo {author} {\bibfnamefont
  {K.}~\bibnamefont {Miuchi}},\ and\ \bibinfo {author} {\bibfnamefont
  {S.}~\bibnamefont {Kurosawa}},\ }\bibfield  {title} {\bibinfo {title} {New
  readout and data-acquisition system in an electron-tracking compton camera
  for {MeV} gamma-ray astronomy ({SMILE-II})},\ }\href
  {https://doi.org/10.1016/j.nima.2015.08.004} {\bibfield  {journal} {\bibinfo
  {journal} {Nuclear Instruments and Methods in Physics Research, Section A:
  Accelerators, Spectrometers, Detectors and Associated Equipment}\ }\textbf
  {\bibinfo {volume} {800}},\ \bibinfo {pages} {40} (\bibinfo {year}
  {2015})}\BibitemShut {NoStop}%
\bibitem [{\citenamefont {Ziegler}\ \emph {et~al.}(2010)\citenamefont
  {Ziegler}, \citenamefont {Ziegler},\ and\ \citenamefont
  {Biersack}}]{Ziegler2010}%
  \BibitemOpen
  \bibfield  {author} {\bibinfo {author} {\bibfnamefont {J.~F.}\ \bibnamefont
  {Ziegler}}, \bibinfo {author} {\bibfnamefont {M.}~\bibnamefont {Ziegler}},\
  and\ \bibinfo {author} {\bibfnamefont {J.}~\bibnamefont {Biersack}},\
  }\bibfield  {title} {\bibinfo {title} {Srim -- the stopping and range of ions
  in matter (2010)},\ }\href
  {https://doi.org/https://doi.org/10.1016/j.nimb.2010.02.091} {\bibfield
  {journal} {\bibinfo  {journal} {Nuclear Instruments and Methods in Physics
  Research Section B: Beam Interactions with Materials and Atoms}\ }\textbf
  {\bibinfo {volume} {268}},\ \bibinfo {pages} {1818} (\bibinfo {year}
  {2010})},\ \bibinfo {note} {19th International Conference on Ion Beam
  Analysis}\BibitemShut {NoStop}%
\bibitem [{Gar()}]{Garfieldpp}%
  \BibitemOpen
  \href@noop {} {}\bibinfo {howpublished}
  {\url{http://garfieldpp.web.cern.ch/garfieldpp/}}\BibitemShut {NoStop}%
\bibitem [{\citenamefont {Biagi}(1999)}]{Biagi1999}%
  \BibitemOpen
  \bibfield  {author} {\bibinfo {author} {\bibfnamefont {S.}~\bibnamefont
  {Biagi}},\ }\bibfield  {title} {\bibinfo {title} {Monte carlo simulation of
  electron drift and diffusion in counting gases under the influence of
  electric and magnetic fields},\ }\href
  {https://doi.org/https://doi.org/10.1016/S0168-9002(98)01233-9} {\bibfield
  {journal} {\bibinfo  {journal} {Nuclear Instruments and Methods in Physics
  Research Section A: Accelerators, Spectrometers, Detectors and Associated
  Equipment}\ }\textbf {\bibinfo {volume} {421}},\ \bibinfo {pages} {234}
  (\bibinfo {year} {1999})}\BibitemShut {NoStop}%
\bibitem [{\citenamefont {Takada}\ \emph {et~al.}(2013)\citenamefont {Takada},
  \citenamefont {Tanimori}, \citenamefont {Kubo}, \citenamefont {Parker},
  \citenamefont {Mizumoto}, \citenamefont {Mizumura}, \citenamefont {Iwaki},
  \citenamefont {Sawano}, \citenamefont {Nakamura}, \citenamefont {Taniue},
  \citenamefont {Higashi}, \citenamefont {Matsuoka}, \citenamefont {Komura},
  \citenamefont {Sato}, \citenamefont {Namamura}, \citenamefont {Oda},
  \citenamefont {Sonoda}, \citenamefont {Tomono}, \citenamefont {Miuchi},
  \citenamefont {Kabuki}, \citenamefont {Kishimoto},\ and\ \citenamefont
  {Kurosawa}}]{Takada2013}%
  \BibitemOpen
  \bibfield  {author} {\bibinfo {author} {\bibfnamefont {A.}~\bibnamefont
  {Takada}}, \bibinfo {author} {\bibfnamefont {T.}~\bibnamefont {Tanimori}},
  \bibinfo {author} {\bibfnamefont {H.}~\bibnamefont {Kubo}}, \bibinfo {author}
  {\bibfnamefont {J.~D.}\ \bibnamefont {Parker}}, \bibinfo {author}
  {\bibfnamefont {T.}~\bibnamefont {Mizumoto}}, \bibinfo {author}
  {\bibfnamefont {Y.}~\bibnamefont {Mizumura}}, \bibinfo {author}
  {\bibfnamefont {S.}~\bibnamefont {Iwaki}}, \bibinfo {author} {\bibfnamefont
  {T.}~\bibnamefont {Sawano}}, \bibinfo {author} {\bibfnamefont
  {K.}~\bibnamefont {Nakamura}}, \bibinfo {author} {\bibfnamefont
  {K.}~\bibnamefont {Taniue}}, \bibinfo {author} {\bibfnamefont
  {N.}~\bibnamefont {Higashi}}, \bibinfo {author} {\bibfnamefont
  {Y.}~\bibnamefont {Matsuoka}}, \bibinfo {author} {\bibfnamefont
  {S.}~\bibnamefont {Komura}}, \bibinfo {author} {\bibfnamefont
  {Y.}~\bibnamefont {Sato}}, \bibinfo {author} {\bibfnamefont {S.}~\bibnamefont
  {Namamura}}, \bibinfo {author} {\bibfnamefont {M.}~\bibnamefont {Oda}},
  \bibinfo {author} {\bibfnamefont {S.}~\bibnamefont {Sonoda}}, \bibinfo
  {author} {\bibfnamefont {D.}~\bibnamefont {Tomono}}, \bibinfo {author}
  {\bibfnamefont {K.}~\bibnamefont {Miuchi}}, \bibinfo {author} {\bibfnamefont
  {S.}~\bibnamefont {Kabuki}}, \bibinfo {author} {\bibfnamefont
  {Y.}~\bibnamefont {Kishimoto}},\ and\ \bibinfo {author} {\bibfnamefont
  {S.}~\bibnamefont {Kurosawa}},\ }\bibfield  {title} {\bibinfo {title}
  {Simulation of gas avalanche in a micro pixel chamber using {Garfield++}},\
  }\href {https://doi.org/10.1088/1748-0221/8/10/C10023} {\bibfield  {journal}
  {\bibinfo  {journal} {Journal of Instrumentation}\ }\textbf {\bibinfo
  {volume} {8}}\bibinfo  {number} { (10)},\ \bibinfo {pages}
  {C10023}}\BibitemShut {NoStop}%
\bibitem [{\citenamefont {Agostinelli}\ \emph {et~al.}(2003)\citenamefont
  {Agostinelli}, \citenamefont {Allison}, \citenamefont {Amako}, \citenamefont
  {Apostolakis}, \citenamefont {Araujo}, \citenamefont {Arce}, \citenamefont
  {Asai}, \citenamefont {Axen}, \citenamefont {Banerjee}, \citenamefont
  {Barrand}, \citenamefont {Behner}, \citenamefont {Bellagamba}, \citenamefont
  {Boudreau}, \citenamefont {Broglia}, \citenamefont {Brunengo}, \citenamefont
  {Burkhardt}, \citenamefont {Chauvie}, \citenamefont {Chuma}, \citenamefont
  {Chytracek}, \citenamefont {Cooperman}, \citenamefont {Cosmo}, \citenamefont
  {Degtyarenko}, \citenamefont {Dell'Acqua}, \citenamefont {Depaola},
  \citenamefont {Dietrich}, \citenamefont {Enami}, \citenamefont {Feliciello},
  \citenamefont {Ferguson}, \citenamefont {Fesefeldt}, \citenamefont {Folger},
  \citenamefont {Foppiano}, \citenamefont {Forti}, \citenamefont {Garelli},
  \citenamefont {Giani}, \citenamefont {Giannitrapani}, \citenamefont {Gibin},
  \citenamefont {{G{\'o}mez Cadenas}}, \citenamefont {Gonz{\'a}lez},
  \citenamefont {{Gracia Abril}}, \citenamefont {Greeniaus}, \citenamefont
  {Greiner}, \citenamefont {Grichine}, \citenamefont {Grossheim}, \citenamefont
  {Guatelli}, \citenamefont {Gumplinger}, \citenamefont {Hamatsu},
  \citenamefont {Hashimoto}, \citenamefont {Hasui}, \citenamefont {Heikkinen},
  \citenamefont {Howard}, \citenamefont {Ivanchenko}, \citenamefont {Johnson},
  \citenamefont {Jones}, \citenamefont {Kallenbach}, \citenamefont {Kanaya},
  \citenamefont {Kawabata}, \citenamefont {Kawabata}, \citenamefont {Kawaguti},
  \citenamefont {Kelner}, \citenamefont {Kent}, \citenamefont {Kimura},
  \citenamefont {Kodama}, \citenamefont {Kokoulin}, \citenamefont {Kossov},
  \citenamefont {Kurashige}, \citenamefont {Lamanna}, \citenamefont
  {Lamp{\'e}n}, \citenamefont {Lara}, \citenamefont {Lefebure}, \citenamefont
  {Lei}, \citenamefont {Liendl}, \citenamefont {Lockman}, \citenamefont
  {Longo}, \citenamefont {Magni}, \citenamefont {Maire}, \citenamefont
  {Medernach}, \citenamefont {Minamimoto}, \citenamefont {{Mora de Freitas}},
  \citenamefont {Morita}, \citenamefont {Murakami}, \citenamefont {Nagamatu},
  \citenamefont {Nartallo}, \citenamefont {Nieminen}, \citenamefont
  {Nishimura}, \citenamefont {Ohtsubo}, \citenamefont {Okamura}, \citenamefont
  {O'Neale}, \citenamefont {Oohata}, \citenamefont {Paech}, \citenamefont
  {Perl}, \citenamefont {Pfeiffer}, \citenamefont {Pia}, \citenamefont
  {Ranjard}, \citenamefont {Rybin}, \citenamefont {Sadilov}, \citenamefont {{Di
  Salvo}}, \citenamefont {Santin}, \citenamefont {Sasaki}, \citenamefont
  {Savvas}, \citenamefont {Sawada}, \citenamefont {Scherer}, \citenamefont
  {Sei}, \citenamefont {Sirotenko}, \citenamefont {Smith}, \citenamefont
  {Starkov}, \citenamefont {Stoecker}, \citenamefont {Sulkimo}, \citenamefont
  {Takahata}, \citenamefont {Tanaka}, \citenamefont {Tcherniaev}, \citenamefont
  {{Safai Tehrani}}, \citenamefont {Tropeano}, \citenamefont {Truscott},
  \citenamefont {Uno}, \citenamefont {Urban}, \citenamefont {Urban},
  \citenamefont {Verderi}, \citenamefont {Walkden}, \citenamefont {Wander},
  \citenamefont {Weber}, \citenamefont {Wellisch}, \citenamefont {Wenaus},
  \citenamefont {Williams}, \citenamefont {Wright}, \citenamefont {Yamada},
  \citenamefont {Yoshida},\ and\ \citenamefont {Zschiesche}}]{Agostinelli2003}%
  \BibitemOpen
\bibfield  {number} {  }\bibfield  {author} {\bibinfo {author} {\bibfnamefont
  {S.}~\bibnamefont {Agostinelli}}, \bibinfo {author} {\bibfnamefont
  {J.}~\bibnamefont {Allison}}, \bibinfo {author} {\bibfnamefont
  {K.}~\bibnamefont {Amako}}, \bibinfo {author} {\bibfnamefont
  {J.}~\bibnamefont {Apostolakis}}, \bibinfo {author} {\bibfnamefont
  {H.}~\bibnamefont {Araujo}}, \bibinfo {author} {\bibfnamefont
  {P.}~\bibnamefont {Arce}}, \bibinfo {author} {\bibfnamefont {M.}~\bibnamefont
  {Asai}}, \bibinfo {author} {\bibfnamefont {D.}~\bibnamefont {Axen}}, \bibinfo
  {author} {\bibfnamefont {S.}~\bibnamefont {Banerjee}}, \bibinfo {author}
  {\bibfnamefont {G.}~\bibnamefont {Barrand}}, \bibinfo {author} {\bibfnamefont
  {F.}~\bibnamefont {Behner}}, \bibinfo {author} {\bibfnamefont
  {L.}~\bibnamefont {Bellagamba}}, \bibinfo {author} {\bibfnamefont
  {J.}~\bibnamefont {Boudreau}}, \bibinfo {author} {\bibfnamefont
  {L.}~\bibnamefont {Broglia}}, \bibinfo {author} {\bibfnamefont
  {A.}~\bibnamefont {Brunengo}}, \bibinfo {author} {\bibfnamefont
  {H.}~\bibnamefont {Burkhardt}}, \bibinfo {author} {\bibfnamefont
  {S.}~\bibnamefont {Chauvie}}, \bibinfo {author} {\bibfnamefont
  {J.}~\bibnamefont {Chuma}}, \bibinfo {author} {\bibfnamefont
  {R.}~\bibnamefont {Chytracek}}, \bibinfo {author} {\bibfnamefont
  {G.}~\bibnamefont {Cooperman}}, \bibinfo {author} {\bibfnamefont
  {G.}~\bibnamefont {Cosmo}}, \bibinfo {author} {\bibfnamefont
  {P.}~\bibnamefont {Degtyarenko}}, \bibinfo {author} {\bibfnamefont
  {A.}~\bibnamefont {Dell'Acqua}}, \bibinfo {author} {\bibfnamefont
  {G.}~\bibnamefont {Depaola}}, \bibinfo {author} {\bibfnamefont
  {D.}~\bibnamefont {Dietrich}}, \bibinfo {author} {\bibfnamefont
  {R.}~\bibnamefont {Enami}}, \bibinfo {author} {\bibfnamefont
  {A.}~\bibnamefont {Feliciello}}, \bibinfo {author} {\bibfnamefont
  {C.}~\bibnamefont {Ferguson}}, \bibinfo {author} {\bibfnamefont
  {H.}~\bibnamefont {Fesefeldt}}, \bibinfo {author} {\bibfnamefont
  {G.}~\bibnamefont {Folger}}, \bibinfo {author} {\bibfnamefont
  {F.}~\bibnamefont {Foppiano}}, \bibinfo {author} {\bibfnamefont
  {A.}~\bibnamefont {Forti}}, \bibinfo {author} {\bibfnamefont
  {S.}~\bibnamefont {Garelli}}, \bibinfo {author} {\bibfnamefont
  {S.}~\bibnamefont {Giani}}, \bibinfo {author} {\bibfnamefont
  {R.}~\bibnamefont {Giannitrapani}}, \bibinfo {author} {\bibfnamefont
  {D.}~\bibnamefont {Gibin}}, \bibinfo {author} {\bibfnamefont
  {J.}~\bibnamefont {{G{\'o}mez Cadenas}}}, \bibinfo {author} {\bibfnamefont
  {I.}~\bibnamefont {Gonz{\'a}lez}}, \bibinfo {author} {\bibfnamefont
  {G.}~\bibnamefont {{Gracia Abril}}}, \bibinfo {author} {\bibfnamefont
  {G.}~\bibnamefont {Greeniaus}}, \bibinfo {author} {\bibfnamefont
  {W.}~\bibnamefont {Greiner}}, \bibinfo {author} {\bibfnamefont
  {V.}~\bibnamefont {Grichine}}, \bibinfo {author} {\bibfnamefont
  {A.}~\bibnamefont {Grossheim}}, \bibinfo {author} {\bibfnamefont
  {S.}~\bibnamefont {Guatelli}}, \bibinfo {author} {\bibfnamefont
  {P.}~\bibnamefont {Gumplinger}}, \bibinfo {author} {\bibfnamefont
  {R.}~\bibnamefont {Hamatsu}}, \bibinfo {author} {\bibfnamefont
  {K.}~\bibnamefont {Hashimoto}}, \bibinfo {author} {\bibfnamefont
  {H.}~\bibnamefont {Hasui}}, \bibinfo {author} {\bibfnamefont
  {A.}~\bibnamefont {Heikkinen}}, \bibinfo {author} {\bibfnamefont
  {A.}~\bibnamefont {Howard}}, \bibinfo {author} {\bibfnamefont
  {V.}~\bibnamefont {Ivanchenko}}, \bibinfo {author} {\bibfnamefont
  {A.}~\bibnamefont {Johnson}}, \bibinfo {author} {\bibfnamefont
  {F.}~\bibnamefont {Jones}}, \bibinfo {author} {\bibfnamefont
  {J.}~\bibnamefont {Kallenbach}}, \bibinfo {author} {\bibfnamefont
  {N.}~\bibnamefont {Kanaya}}, \bibinfo {author} {\bibfnamefont
  {M.}~\bibnamefont {Kawabata}}, \bibinfo {author} {\bibfnamefont
  {Y.}~\bibnamefont {Kawabata}}, \bibinfo {author} {\bibfnamefont
  {M.}~\bibnamefont {Kawaguti}}, \bibinfo {author} {\bibfnamefont
  {S.}~\bibnamefont {Kelner}}, \bibinfo {author} {\bibfnamefont
  {P.}~\bibnamefont {Kent}}, \bibinfo {author} {\bibfnamefont {A.}~\bibnamefont
  {Kimura}}, \bibinfo {author} {\bibfnamefont {T.}~\bibnamefont {Kodama}},
  \bibinfo {author} {\bibfnamefont {R.}~\bibnamefont {Kokoulin}}, \bibinfo
  {author} {\bibfnamefont {M.}~\bibnamefont {Kossov}}, \bibinfo {author}
  {\bibfnamefont {H.}~\bibnamefont {Kurashige}}, \bibinfo {author}
  {\bibfnamefont {E.}~\bibnamefont {Lamanna}}, \bibinfo {author} {\bibfnamefont
  {T.}~\bibnamefont {Lamp{\'e}n}}, \bibinfo {author} {\bibfnamefont
  {V.}~\bibnamefont {Lara}}, \bibinfo {author} {\bibfnamefont {V.}~\bibnamefont
  {Lefebure}}, \bibinfo {author} {\bibfnamefont {F.}~\bibnamefont {Lei}},
  \bibinfo {author} {\bibfnamefont {M.}~\bibnamefont {Liendl}}, \bibinfo
  {author} {\bibfnamefont {W.}~\bibnamefont {Lockman}}, \bibinfo {author}
  {\bibfnamefont {F.}~\bibnamefont {Longo}}, \bibinfo {author} {\bibfnamefont
  {S.}~\bibnamefont {Magni}}, \bibinfo {author} {\bibfnamefont
  {M.}~\bibnamefont {Maire}}, \bibinfo {author} {\bibfnamefont
  {E.}~\bibnamefont {Medernach}}, \bibinfo {author} {\bibfnamefont
  {K.}~\bibnamefont {Minamimoto}}, \bibinfo {author} {\bibfnamefont
  {P.}~\bibnamefont {{Mora de Freitas}}}, \bibinfo {author} {\bibfnamefont
  {Y.}~\bibnamefont {Morita}}, \bibinfo {author} {\bibfnamefont
  {K.}~\bibnamefont {Murakami}}, \bibinfo {author} {\bibfnamefont
  {M.}~\bibnamefont {Nagamatu}}, \bibinfo {author} {\bibfnamefont
  {R.}~\bibnamefont {Nartallo}}, \bibinfo {author} {\bibfnamefont
  {P.}~\bibnamefont {Nieminen}}, \bibinfo {author} {\bibfnamefont
  {T.}~\bibnamefont {Nishimura}}, \bibinfo {author} {\bibfnamefont
  {K.}~\bibnamefont {Ohtsubo}}, \bibinfo {author} {\bibfnamefont
  {M.}~\bibnamefont {Okamura}}, \bibinfo {author} {\bibfnamefont
  {S.}~\bibnamefont {O'Neale}}, \bibinfo {author} {\bibfnamefont
  {Y.}~\bibnamefont {Oohata}}, \bibinfo {author} {\bibfnamefont
  {K.}~\bibnamefont {Paech}}, \bibinfo {author} {\bibfnamefont
  {J.}~\bibnamefont {Perl}}, \bibinfo {author} {\bibfnamefont {A.}~\bibnamefont
  {Pfeiffer}}, \bibinfo {author} {\bibfnamefont {M.}~\bibnamefont {Pia}},
  \bibinfo {author} {\bibfnamefont {F.}~\bibnamefont {Ranjard}}, \bibinfo
  {author} {\bibfnamefont {A.}~\bibnamefont {Rybin}}, \bibinfo {author}
  {\bibfnamefont {S.}~\bibnamefont {Sadilov}}, \bibinfo {author} {\bibfnamefont
  {E.}~\bibnamefont {{Di Salvo}}}, \bibinfo {author} {\bibfnamefont
  {G.}~\bibnamefont {Santin}}, \bibinfo {author} {\bibfnamefont
  {T.}~\bibnamefont {Sasaki}}, \bibinfo {author} {\bibfnamefont
  {N.}~\bibnamefont {Savvas}}, \bibinfo {author} {\bibfnamefont
  {Y.}~\bibnamefont {Sawada}}, \bibinfo {author} {\bibfnamefont
  {S.}~\bibnamefont {Scherer}}, \bibinfo {author} {\bibfnamefont
  {S.}~\bibnamefont {Sei}}, \bibinfo {author} {\bibfnamefont {V.}~\bibnamefont
  {Sirotenko}}, \bibinfo {author} {\bibfnamefont {D.}~\bibnamefont {Smith}},
  \bibinfo {author} {\bibfnamefont {N.}~\bibnamefont {Starkov}}, \bibinfo
  {author} {\bibfnamefont {H.}~\bibnamefont {Stoecker}}, \bibinfo {author}
  {\bibfnamefont {J.}~\bibnamefont {Sulkimo}}, \bibinfo {author} {\bibfnamefont
  {M.}~\bibnamefont {Takahata}}, \bibinfo {author} {\bibfnamefont
  {S.}~\bibnamefont {Tanaka}}, \bibinfo {author} {\bibfnamefont
  {E.}~\bibnamefont {Tcherniaev}}, \bibinfo {author} {\bibfnamefont
  {E.}~\bibnamefont {{Safai Tehrani}}}, \bibinfo {author} {\bibfnamefont
  {M.}~\bibnamefont {Tropeano}}, \bibinfo {author} {\bibfnamefont
  {P.}~\bibnamefont {Truscott}}, \bibinfo {author} {\bibfnamefont
  {H.}~\bibnamefont {Uno}}, \bibinfo {author} {\bibfnamefont {L.}~\bibnamefont
  {Urban}}, \bibinfo {author} {\bibfnamefont {P.}~\bibnamefont {Urban}},
  \bibinfo {author} {\bibfnamefont {M.}~\bibnamefont {Verderi}}, \bibinfo
  {author} {\bibfnamefont {A.}~\bibnamefont {Walkden}}, \bibinfo {author}
  {\bibfnamefont {W.}~\bibnamefont {Wander}}, \bibinfo {author} {\bibfnamefont
  {H.}~\bibnamefont {Weber}}, \bibinfo {author} {\bibfnamefont
  {J.}~\bibnamefont {Wellisch}}, \bibinfo {author} {\bibfnamefont
  {T.}~\bibnamefont {Wenaus}}, \bibinfo {author} {\bibfnamefont
  {D.}~\bibnamefont {Williams}}, \bibinfo {author} {\bibfnamefont
  {D.}~\bibnamefont {Wright}}, \bibinfo {author} {\bibfnamefont
  {T.}~\bibnamefont {Yamada}}, \bibinfo {author} {\bibfnamefont
  {H.}~\bibnamefont {Yoshida}},\ and\ \bibinfo {author} {\bibfnamefont
  {D.}~\bibnamefont {Zschiesche}},\ }\bibfield  {title} {\bibinfo {title}
  {Geant4---a simulation toolkit},\ }\href
  {https://doi.org/https://doi.org/10.1016/S0168-9002(03)01368-8} {\bibfield
  {journal} {\bibinfo  {journal} {Nuclear Instruments and Methods in Physics
  Research Section A: Accelerators, Spectrometers, Detectors and Associated
  Equipment}\ }\textbf {\bibinfo {volume} {506}},\ \bibinfo {pages} {250}
  (\bibinfo {year} {2003})}\BibitemShut {NoStop}%
\bibitem [{\citenamefont {Utsunomiya}\ \emph {et~al.}(2018)\citenamefont
  {Utsunomiya}, \citenamefont {Watanabe}, \citenamefont {Ari-izumi},
  \citenamefont {Takenaka}, \citenamefont {Araki}, \citenamefont {Tsuji},
  \citenamefont {Gheorghe}, \citenamefont {Filipescu}, \citenamefont
  {Belyshev}, \citenamefont {Stopani}, \citenamefont {Symochko}, \citenamefont
  {Wang}, \citenamefont {Fan}, \citenamefont {Renstr{\o}m}, \citenamefont
  {Tveten}, \citenamefont {Lui}, \citenamefont {Sugita},\ and\ \citenamefont
  {Miyamoto}}]{Utsunomiya2018}%
  \BibitemOpen
  \bibfield  {author} {\bibinfo {author} {\bibfnamefont {H.}~\bibnamefont
  {Utsunomiya}}, \bibinfo {author} {\bibfnamefont {T.}~\bibnamefont
  {Watanabe}}, \bibinfo {author} {\bibfnamefont {T.}~\bibnamefont {Ari-izumi}},
  \bibinfo {author} {\bibfnamefont {D.}~\bibnamefont {Takenaka}}, \bibinfo
  {author} {\bibfnamefont {T.}~\bibnamefont {Araki}}, \bibinfo {author}
  {\bibfnamefont {K.}~\bibnamefont {Tsuji}}, \bibinfo {author} {\bibfnamefont
  {I.}~\bibnamefont {Gheorghe}}, \bibinfo {author} {\bibfnamefont {D.~M.}\
  \bibnamefont {Filipescu}}, \bibinfo {author} {\bibfnamefont {S.}~\bibnamefont
  {Belyshev}}, \bibinfo {author} {\bibfnamefont {K.}~\bibnamefont {Stopani}},
  \bibinfo {author} {\bibfnamefont {D.}~\bibnamefont {Symochko}}, \bibinfo
  {author} {\bibfnamefont {H.}~\bibnamefont {Wang}}, \bibinfo {author}
  {\bibfnamefont {G.}~\bibnamefont {Fan}}, \bibinfo {author} {\bibfnamefont
  {T.}~\bibnamefont {Renstr{\o}m}}, \bibinfo {author} {\bibfnamefont {G.~M.}\
  \bibnamefont {Tveten}}, \bibinfo {author} {\bibfnamefont {Y.-W.}\
  \bibnamefont {Lui}}, \bibinfo {author} {\bibfnamefont {K.}~\bibnamefont
  {Sugita}},\ and\ \bibinfo {author} {\bibfnamefont {S.}~\bibnamefont
  {Miyamoto}},\ }\bibfield  {title} {\bibinfo {title} {Photon-flux
  determination by the poisson-fitting technique with quenching corrections},\
  }\href {https://doi.org/https://doi.org/10.1016/j.nima.2018.04.021}
  {\bibfield  {journal} {\bibinfo  {journal} {Nuclear Instruments and Methods
  in Physics Research Section A: Accelerators, Spectrometers, Detectors and
  Associated Equipment}\ }\textbf {\bibinfo {volume} {896}},\ \bibinfo {pages}
  {103} (\bibinfo {year} {2018})}\BibitemShut {NoStop}%
\bibitem [{\citenamefont {Furuno}\ \emph {et~al.}(2019)\citenamefont {Furuno},
  \citenamefont {Kawabata}, \citenamefont {Adachi}, \citenamefont {Ayyad},
  \citenamefont {Kanada-En'yo}, \citenamefont {Fujikawa}, \citenamefont
  {Inaba}, \citenamefont {Murata}, \citenamefont {Ong}, \citenamefont
  {Sferrazza}, \citenamefont {Takahashi}, \citenamefont {Takeda}, \citenamefont
  {Tanihata}, \citenamefont {Tran},\ and\ \citenamefont
  {Tsumura}}]{Furuno2019}%
  \BibitemOpen
  \bibfield  {author} {\bibinfo {author} {\bibfnamefont {T.}~\bibnamefont
  {Furuno}}, \bibinfo {author} {\bibfnamefont {T.}~\bibnamefont {Kawabata}},
  \bibinfo {author} {\bibfnamefont {S.}~\bibnamefont {Adachi}}, \bibinfo
  {author} {\bibfnamefont {Y.}~\bibnamefont {Ayyad}}, \bibinfo {author}
  {\bibfnamefont {Y.}~\bibnamefont {Kanada-En'yo}}, \bibinfo {author}
  {\bibfnamefont {Y.}~\bibnamefont {Fujikawa}}, \bibinfo {author}
  {\bibfnamefont {K.}~\bibnamefont {Inaba}}, \bibinfo {author} {\bibfnamefont
  {M.}~\bibnamefont {Murata}}, \bibinfo {author} {\bibfnamefont {H.~J.}\
  \bibnamefont {Ong}}, \bibinfo {author} {\bibfnamefont {M.}~\bibnamefont
  {Sferrazza}}, \bibinfo {author} {\bibfnamefont {Y.}~\bibnamefont
  {Takahashi}}, \bibinfo {author} {\bibfnamefont {T.}~\bibnamefont {Takeda}},
  \bibinfo {author} {\bibfnamefont {I.}~\bibnamefont {Tanihata}}, \bibinfo
  {author} {\bibfnamefont {D.~T.}\ \bibnamefont {Tran}},\ and\ \bibinfo
  {author} {\bibfnamefont {M.}~\bibnamefont {Tsumura}},\ }\bibfield  {title}
  {\bibinfo {title} {Neutron quadrupole transition strength in
  $^{10}\mathrm{C}$ deduced from the
  $^{10}\mathrm{C}(\ensuremath{\alpha},{\ensuremath{\alpha}}^{\ensuremath{'}})$
  measurement with the {MAIKo} active target},\ }\href
  {https://doi.org/10.1103/PhysRevC.100.054322} {\bibfield  {journal} {\bibinfo
   {journal} {Phys. Rev. C}\ }\textbf {\bibinfo {volume} {100}},\ \bibinfo
  {pages} {054322} (\bibinfo {year} {2019})}\BibitemShut {NoStop}%
\end{thebibliography}%

\end{document}